\documentclass{aa}  

\usepackage{graphicx}
\usepackage{booktabs}
\usepackage{supertabular}
\usepackage[version=4]{mhchem}
\usepackage{ulem}
\usepackage{xcolor}
\usepackage{amsmath}
\usepackage{txfonts}

\begin{document}

   \title{Importance of source structure on complex organics emission}

   \subtitle{II. Can disks explain lack of methanol emission from some low-mass protostars?}

   \author{P. Nazari
          \inst{1}
          \and
          B. Tabone\inst{1}
          \and
          G. P. Rosotti\inst{1,2}
          \and
          M. L. van Gelder\inst{1}
          \and
          R. Meshaka\inst{3}
          \and
          E. F. van Dishoeck\inst{1,4}
          }

   \institute{Leiden Observatory, Leiden University, P.O. Box 9513, 2300 RA Leiden, the Netherlands\\ 
        \email{nazari@strw.leidenuniv.nl}
         \and
          School of Physics and Astronomy, University of Leicester, Leicester LE1 7RH, UK
         \and
         LERMA, Observatoire de Paris, Université PSL, CNRS, Sorbonne Université, F-92195 Meudon, France
         \and
        Max Planck Institut f\"{u}r Extraterrestrische Physik (MPE), Giessenbachstrasse 1, 85748 Garching, Germany
             }

   \date{Received XXX; accepted YYY}

 
  \abstract
   {The protostellar stage is known to be the richest star formation phase in emission from gaseous complex organic molecules. However, some protostellar systems show little or no millimetre (mm) line emission of such species. This can be interpreted as a low abundance of complex organic molecules, alternatively complex species could be present in the system but are not seen in the gas.}
   {The goal is to investigate the latter hypothesis for methanol as the most abundant complex organic molecule in protostellar systems. This work aims to find out how effective dust optical depth is in hiding methanol in the gas and whether methanol can mainly reside in the ice due to the presence of a disk that lowers the temperatures. Hence, we will attempt to answer the question: Does the presence of a disk and optically thick dust reduce methanol emission even if methanol and other complex species are abundant in the ices and gas?}
   {Using the radiative transfer code RADMC-3D, methanol emission lines from an envelope-only model and an envelope-plus-disk model are calculated and compared with each other and the observations. Methanol gas and ice abundances are parameterised inside and outside of the snow surfaces based on values from observations. Both models include either dust grains with low mm opacity or high mm opacity and their physical parameters such as envelope mass and disk radius are varied.}
   {Methanol emission from the envelope-only model is always stronger than from the envelope-plus-disk model by at least a factor ${\sim} 2$ as long as the disk radius is larger than ${\sim} 30$\,au (for $L=8$\,L$_{\odot}$). In most cases, this is due to lower temperatures (disk shadowing) and, hence, the smaller amount of warm ($\gtrsim 70$\,K) methanol inside the snow surface of the envelope-plus-disk model. The intensities drop by more than an order of magnitude for models including high mm opacity dust grains and disk radii of at least ${\sim} 50$\,au (for $L=8$\,L$_{\odot}$) due to continuum over-subtraction.}
   {The line intensities from the envelope-only models match the observations moderately well when methanol emission is strong but overproduce the observations of protostars with lower methanol emission even with large dust optical depth effects. The envelope-plus-disk models can explain the bulk of the observations. However, they can only reproduce the observations of sources with high luminosities and very low methanol emission when dust optical depth is significant in the envelope and continuum over-subtraction becomes effective in the disk (high mm opacity dust grains are used). Therefore, both the effects of disk and dust optical depth should be considered to explain the observations. In conclusion, it is important to take physical structure into account in future chemical studies of low-mass protostars: Absence of gas-phase methanol emission does not imply absence of methanol molecules in either gas or ice.}


   \maketitle
%

\section{Introduction}

The protostellar phase of star formation is the stage at which most of the material surrounding the protostar is warm and hence species existing in ices can sublimate into the gas phase and result in this stage becoming the most line rich one (\citealt{Herbst2009}; \citealt{Caselli2012}; \citealt{vantHoff2020c}). Therefore, the earlier stages of star formation provide the best opportunity to study complex organics in the gas phase. 

Methanol (CH$_3$OH) is the most common complex organic molecule detected towards both low- and high-mass protostars in the gas over the past decades (e.g., \citealt{Blake1987}; \citealt{Ewine1995}; \citealt{Gibb2000}; \citealt{Bisschop2007}; \citealt{Belloche2013}; \citealt{Jorgensen2016}; \citealt{Ilee2016}; \citealt{Bogelund2018}; \citealt{Marcelino2018}; \citealt{Martin-Domenech2019}; \citealt{Taquet2019}; \citealt{vangelder2020}; \citealt{Manigand2020}; \citealt{Yang2020}; \citealt{Ligterink2021}; \citealt{Law2021}). Moreover, methanol has been observed to be abundant ($X_{\rm{H}} {\sim}3 \times 10^{-6}$, with some spread) in ices towards protostellar objects (\citealt{Geballe1988}; \citealt{Dartois1999}; \citealt{Dartois2002}; \citealt{Boogert2008}; \citealt{Bottinelli2010}; \citealt{Oberg2011}; \citealt{Boogert2015}) and in some dark cores prior to star formation (\citealt{Boogert2011}; Qasim, D.N. thesis 2020). A high methanol abundance in both gas and solids is only possible if methanol is formed in the ice as observed in the lab experiments and chemical models (\citealt{Hidaka2004}; \citealt{Geppert2006}; \citealt{Fuchs2009}; \citealt{Garrod2011}). 

However, some sources do not show methanol emission at millimetre wavelengths in the gas which has become more apparent from the recent Atacama Large Millimeter/submillimeter Array (ALMA) and the Northern Extended Millimeter Array (NOEMA) surveys. \cite{Yang2021} detect methanol in 56\% of their 50 low-mass protostars and \cite{Belloche2020} detect methanol in 50\% of the 26 low-mass protostars they observe.

This has been investigated further in \cite{vanGelder2022} who find that there are four orders of magnitude spread in the mass of gaseous methanol across protostars of similar luminosity that have been observed by ALMA. Does this mean that the amount of methanol strongly varies in the hot cores? Modelling work of \cite{Drozdovskaya2014} shows that the abundance of methanol can increase or decrease through infall going from the pre-stellar core to the protoplanetary disk depending on the disk growth mechanisms. Hence, the spread in column density of methanol as found by \cite{vanGelder2022} could be partially explained by the loss of methanol during the infall or the destruction of methanol through some other process (dehydrogenation, when a molecule loses a hydrogen atom through some chemical reaction; \citealt{Nourry2015}; photodissociation; \citealt{Laas2011}; \citealt{McGuire2017}; \citealt{Notsu2021}) resulting in simply a lower methanol number density in some protostellar systems. However, \cite{Aikawa2020} show with their chemical models that even if pristine methanol is destroyed at the end of the dark cloud phase, it will be efficiently reformed in the ice via reaction of CH$_3$ and OH once the collapse starts. Therefore, the possibility of the absence of methanol in protostellar systems is less likely and hence there is a necessity for another explanation. 

There is a similar debate for water abundances in protostellar systems (\citealt{Persson2012,Persson2016}; \citealt{Visser2013}; \citealt{vanDishoeck2021}) which seem to be orders of magnitude lower than that expected (${\sim} 10^{-4}$) from ice sublimation. Several explanations are proposed for this discrepancy based on chemistry and physical structures of the regions in the literature. \cite{Notsu2021} find that X-ray induced chemistry of water and other molecules such as methanol decreases the abundances of these molecules in the inner protostellar region. Moreover, \cite{Persson2016} show that the presence of a disk significantly reduces the amount of gas and dust above 100\,K, potentially lowering abundances of water by an order of magnitude. The same argument can be applied to methanol as it has a similar sublimation temperature as water (${\sim} 100$\,K). 

Thus, it is possible that methanol is not observed in the gas because this molecule resides in the ice. The fact that a disk can decrease the temperature has been discussed in the literature (\citealt{Persson2016}; \citealt{Murillo2015}). In particular, \cite{Murillo2018} observed DCO$^{+}$, a cold gas tracer (D/H is enhanced at lower temperatures), behind the location of the disk in IRAS 16293-2422A and VLA 1623-2417. They conclude that the lower temperatures behind the disk are caused by disk shadowing and produce an environment favourable for the formation of deuterated species such as DCO$^{+}$.

Another reason for methanol not being observed in the gas can be dust attenuation blocking the methanol emission. This has been directly observed by \cite{DeSimone2020}, for the case of NGC 1333 IRAS 4A1, where they detect methanol at centimetre wavelengths while the same source had not shown any complex organic emission at millimetre wavelengths (\citealt{Lopez-spulcre2017}). Therefore, sources which appear poor in complex organic molecules with ALMA may in fact be rich in gaseous species but hidden at ALMA wavelengths.

In this work, we investigate the scenario in which methanol is plentiful in the protostellar systems but cannot be seen in the gas phase: Could the drop in temperature as a result of the presence of a disk in addition to dust optical depth explain the lack of gaseous methanol emission in some sources? This is done by comparing the results of two different models: an envelope-only model and a flattened envelope-plus-disk model. In both cases, we make sure to consider optically thin and thick millimetre continuum to examine the effects of dust attenuation as observed by \cite{DeSimone2020}. Methanol ice and gas abundances are simply parameterised in the models based on the values from observations and inspired by detailed chemical disk models rather than being calculated by including a complete chemical network.

In Sect. \ref{sec:methods} the models and the assumptions made are explained, Sect. \ref{sec:results} presents the results especially on the thermal structure and the methanol emission from the models. In Sect. \ref{sec:discussion} we discuss our findings and compare our results with observations of protostars with ALMA. Finally Sect. \ref{sec:conslusion} presents our conclusions.

\section{Methods}
\label{sec:methods}
\subsection{Physical structure}

\subsubsection{Envelope-only model}
\label{sec:env_model}

\begin{figure*}
    \centering
    \includegraphics[width=0.8\textwidth]{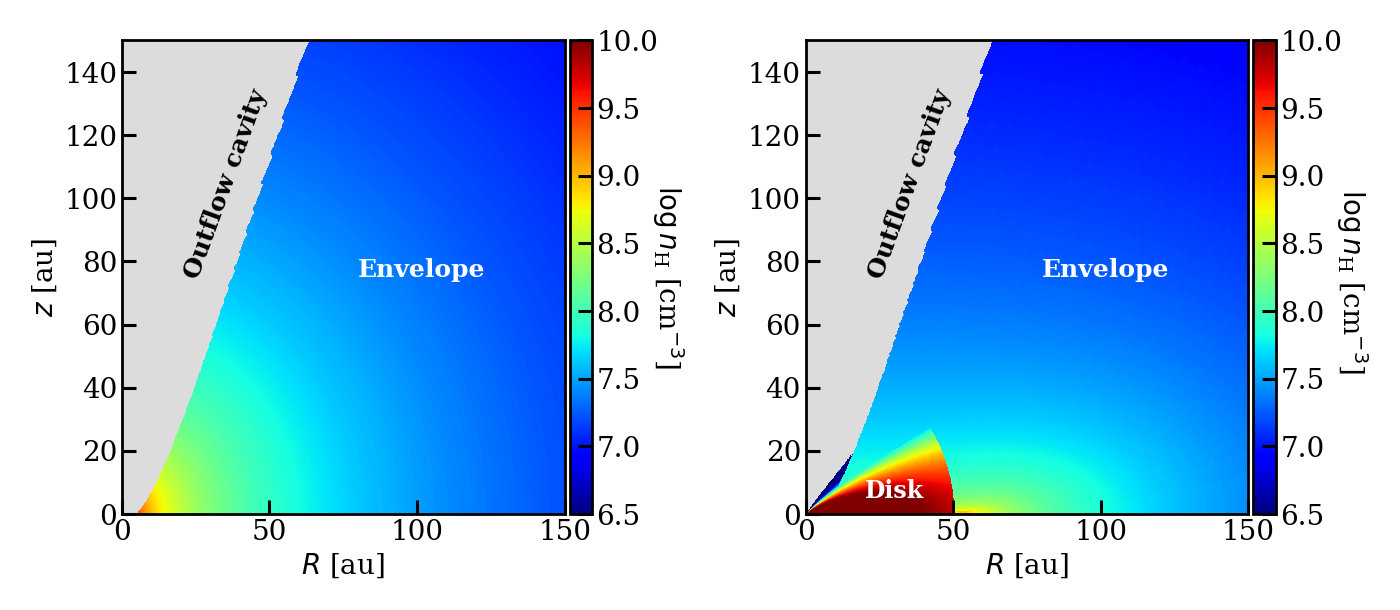}
    \caption{Density profiles explored in this work. Left panel: 2D hydrogen nuclei number density, $n_{\rm H} = n(\textup{H}) + 2 n(\rm{H}_2)$, for the fiducial envelope-only model. Right panel: the same as left panel but for the fiducial flattened envelope-plus-disk model. In this figure and all the other figures, $\log$ is in base 10.} 
    \label{fig:H_density}
\end{figure*}

The left panel of Fig. \ref{fig:H_density} shows the density structure for our fiducial envelope-only protostellar system with envelope mass of 1\,M$_\odot$ and protostellar luminosity of 8\,L$_{\odot}$ (values similar to those for low-mass protostars rich in complex organics, e.g. L1448 IRS2 and L1448 IRS3; \citealt{Mottram2017}). The spherically symmetric envelope model assumes a power-law gas density relation with radius,

\begin{equation}
    \rho_{\rm g} = \rho_{0} \left(\frac{r}{r_{0}}\right)^{-\alpha}, 
    \label{eq:density_env}
\end{equation}

\noindent where $\rho_{\rm g}$ is the gas density of the envelope, $r$ is the radius in spherical coordinates, $\rho_{0}$ is the gas density at radius $r_{0}$ which is parameterised by the envelope mass and assuming inner and outer radii of the envelope to be 0.4\,au and 10$^4$\,au. $\alpha$ is fixed to 1.5 to represent the values found from observations of protostellar envelopes on scales of ${\sim}500$\,au and larger (\citealt{Kristensen2012}). A gas-to-dust mass ratio of 100 is assumed to calculate the dust density. 

An outflow cavity is also added to the model with the same shape as the outflow cavity in the envelope-plus-disk model explained in Sect. \ref{sec:env+disk_model}. The density inside the cavity is fixed to $10^{3}\, \rm cm^{-3}$ to be in line with the observations (\citealt{Bachiller1999}; \citealt{Whitney2003}).

A 5000\,K blackbody is assumed for the input radiation field to mimic the effect of accretion luminosity and protostellar radiation field. The envelope mass is varied from $0.1\,\rm M_{\odot}$ to $10\,\rm M_{\odot}$ and the luminosity from $0.5\,\rm L_{\odot}$ to $32\,\rm L_{\odot}$. The envelope masses and luminosities are chosen so that they represent well-studied Class 0 objects (\citealt{Jorgensen2009}; \citealt{Kristensen2012}). Table \ref{tab:params} summarises the parameters used for the envelope-only model.

\begin{table*}[t]
  \centering
  \caption{Parameters of the models}
  \label{tab:params}
  \begin{tabular}{lllll}
  \toprule
  \toprule
  Parameter [unit] & Envelope-only  & Envelope-plus-disk & Description  \\
  \midrule
    $r_{\rm in}$ [au] & 0.4 & 0.4 & The inner radius\\
    $r_{\rm out}$ [au]  & $10^{4}$ & $10^{4}$ & The outer radius of the envelope\\
    $M_{\rm E}$ [M$_{\odot}$] & 0.1, 0.3, \textbf{1}, 3, & 0.1, 0.3, \textbf{1}, 3, & Envelope mass\\& 5, 10 & 5, 10 & \\
    $R_{\rm D}$ [au] & -- & 10, 20, 30, \textbf{50}, & Disk radius\\&& 100, 200 & \\
    $T_{\star}$ [K]  & 5000 & 5000 & Protostellar temperature\\
    $M_{\star}$[M$_{\odot}$]  & 0.5 & 0.5 & Protostellar mass\\
    $L$ [L$_{\odot}$] & 0.5, 1, 2, 4 , \textbf{8}, & 0.5, 1, 2, 4 , \textbf{8}, & Bolometric luminosity\\& 16, 32 & 16, 32 & \\
    \bottomrule
  \end{tabular}
  \tablefoot{The parameters of the fiducial model are highlighted with bold face.}
\end{table*}

\subsubsection{Envelope-plus-disk model}
\label{sec:env+disk_model}
\begin{figure}
  \resizebox{0.9\hsize}{!}{\includegraphics{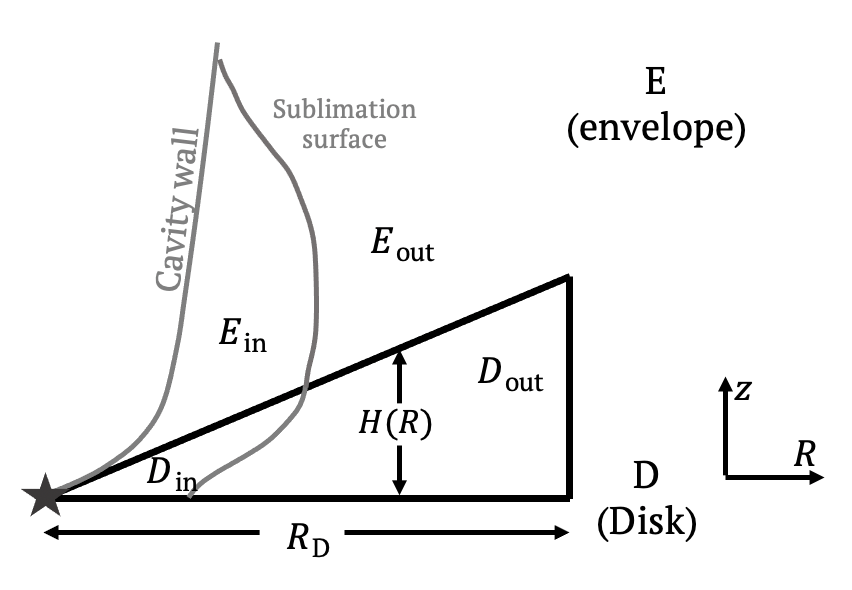}}
  \caption{A cartoon of the envelope-plus-disk model's physical structure. $H(R)$ and $R_{\rm D}$ are the disk scale height and the disk radius. $D_{\rm in}$ and $D_{\rm out}$ are the inner and outer part of the disk with respect to the sublimation surface. $E_{\rm in}$ and $E_{\rm out}$ are the same but for the envelope.}
  \label{fig:cartoon_structure}
\end{figure}

A disk is included through a parameterised disk plus a flattened envelope model following the works by \cite{Crapsi2008} and \cite{Harsono2015}.  A sketch of the envelope-plus-disk physical structure is shown in Fig. \ref{fig:cartoon_structure} and the density structure for our fiducial envelope-plus-disk protostellar system with envelope mass of 1\,M$_\odot$, stellar luminosity of 8\,L$_\odot$, disk radius of 50\,au and disk mass of 0.01\,M$_\odot$ is shown in the right panel of Fig. \ref{fig:H_density}. The disk mass and radius for the fiducial envelope-plus-disk model are chosen so that they agree with the values for typical Class 0 sources (\citealt{Murillo2013}; \citealt{Maury2019}; \citealt{Tobin2020}).

The gas density of the flattened envelope following \cite{Ulrich1976} is  

\begin{align}
    \label{eq:dens}
    \begin{split}
      \rho_{\rm g,E}  \propto & ~ \left(\frac{r}{r_{\rm out}}\right)^{-3/2} (1+\cos{\theta}/\cos{\theta_0})^{-1/2}\\& 
     \times \left(\frac{\cos{\theta}}{2\cos{\theta_0}}+\frac{r_{\rm c}}{r} \cos^2{\theta_0}\right)^{-1},
     \end{split}
\end{align}

\noindent where $r$ is the radius in spherical coordinates, $r_{\rm out}$ is the outer radius, $r_{\rm c}$ is the centrifugal radius which has been assumed to be 50\,au, and $\theta_{0}$ is the initial latitude of each particle. The term, $\cos{\theta_{0}}$, is calculated for each point of the grid by solving the following equation

\begin{equation}
    \frac{r}{r_{\rm c}}\left(1- \cos{\theta}/\cos{\theta_0} \right) = 1-\cos^2{\theta_0}.
    \label{eq:cos_theta0}
\end{equation}

An outflow cavity is added to the model in a similar way to the envelope-only model: by fixing the hydrogen density to $10^3 \, \rm cm^{-3}$ where $\cos{\theta_0}$ from Eq. \eqref{eq:cos_theta0} is higher than 0.95 to be consistent with the observations and modelling of outflows (\citealt{Crapsi2008}; \citealt{Plunkett2013}; \citealt{Harsono2015}). 

A disk is added to the flattened envelope with power-law gas density in radius and Gaussian distribution in $z$ direction in cylindrical coordinates expected from hydrostatic equilibrium. The density is given by (\citealt{Shakura1973}; \citealt{Pringle1981})  

\begin{equation}
    \rho_{\rm g,D} = \frac{M_{\rm D} (R/R_{\rm D})^{-1}}{\sqrt{8 \pi^{3}} H(R) R_{\rm D}^{2}} \exp\left[-\frac{1}{2}\left(\frac{z}{H(R)}\right)^{2}\right],
\end{equation}

\noindent where $M_{\rm D}$ is the disk mass, $R_{\rm D}$ is the disk radius, $z$ is the height from the disk mid-plane, $R$ is the radius in cylindrical coordinates and $H(R)$ is the scale height of the disk that is dependent on the radius. The disk radius, $R_{\rm D}$, is the disk's physical size and is different from the gas and dust disk sizes measured from the observations, with gas sizes generally larger than those of the dust. $H(R)$ is set by the thermal structure of the disk. The disk density is assumed to be zero at radii larger than the disk radius. Assuming a vertically isothermal disk the scale height is given by $\epsilon R$, where $\epsilon$ is the aspect ratio of the disk and is fixed to 0.2 (\citealt{Chiang1997}).

The total gas density in the envelope-plus-disk model is given by

\begin{equation}
\rho_{g,D+E} = 
\begin{cases}
    \rho_{g,D},& \rho_{g,D} > \rho_{g,E},\\
    \rho_{g,E},& \rho_{g,E} \geq \rho_{g,D}.
\end{cases}
\label{eq:rho_disk_final}
\end{equation}

\noindent In Eq. \eqref{eq:rho_disk_final}, the outflow cavity and the envelope are assumed to be part of the same component represented by $\rho_{g,E}$. The dust density is calculated by assuming a gas-to-dust mass ratio of 100.

The most important parameters of the disk are its mass and radius that are varied as shown in Table \ref{tab:params}. The disk mass and radius are varied simultaneously to keep $M_{\rm D}/R_{\rm D}^{2}$ (an approximation to the disk surface density) constant in all models. The envelope mass and protostellar luminosity are varied in the same way as the envelope-only models (see Sect. \ref{sec:env_model}).

\subsection{Methanol abundance}
\label{sec:meth_abund}

The aim of this work is to calculate the emission from gas-phase methanol in protostellar systems. Therefore, it is crucial to have a realistic model for the methanol abundance. This is parameterised in the models based on the various regions of the disk and envelope. These regions are marked on Fig. \ref{fig:cartoon_structure}: $E_{\rm in}$ and $E_{\rm out}$ are the inner and outer envelope regions, $D_{\rm in}$ and $D_{\rm out}$ are the inner and outer disk regions where inner means inside of the sublimation surface and outer means outside of it.  

We use the balance between adsorption and thermal desorption of methanol (\citealt{Hasegawa1992}) to calculate the snow surface and hence abundance ($X$) with respect to $n_{\rm H} = n(\textup{H}) + 2 n(\rm{H}_2)$ at each point in the envelope-only model and the envelope-plus-disk model. The ice-to-gas abundance ratio of methanol ($X_{\rm ice}/ X_{\rm gas}$) is equal to the ratio of number density of the molecules in the solid phase to the gas phase that is given by

\begin{equation}
    \frac{X_{\rm ice}}{X_{\rm gas}} = \frac{n_{\rm ice}}{n_{\rm gas}} = \frac{\pi a_{\rm d}^{2} n_{\rm d} S \sqrt{3k_{\rm B}T_{\rm gas}/m_{i}}}{e^{-E_{\rm b}/T_{\rm d}} \sqrt{2 k_{\rm B} n_{\rm ss} E_{\rm b}/(\pi^{2}m_{i})}},
    \label{eq:gas-grain}
\end{equation}

\noindent where $a_{\rm d}$ is the characteristic dust grain size (see Appendix \ref{app:meth_abund}), $n_{\rm d}$ is the dust number density, $S$ is the sticking coefficient which is assumed to be 1, $k_{\rm B}$ is the Boltzmann constant, $T_{\rm gas}$ is the gas temperature, $T_{\rm d}$ is the dust temperature assumed to be the same as $T_{\rm gas}$ due to thermal coupling of the gas and dust in the dense regions, $m_{i}$ is the mass of the species $i$, $E_{\rm b}$ is the binding energy assumed to be 3820\,K for methanol (\citealt{Penteado2017}) and $n_{\rm ss}$ is the number of binding sites per surface area assumed to be $8 \times 10^{14} \rm cm^{-2}$ (\citealt{Visser2011}). These values result in a desorption temperature of ${\sim}70$\,K for the typical densities of our models ($n_{\rm H} {\sim} 10^{7}$\,cm$^{-3}$). 

A total methanol abundance ($X_{\rm ice} + X_{\rm gas}$) of $10^{-6}$ is assumed in the envelope-only model and the envelope component of the envelope-plus-disk model. Moreover, where $X_{\rm gas}$ goes below $10^{-9}$ the gas-phase abundance is set to $10^{-9}$ (i.e. abundance in $E_{\rm out}$ in Fig. \ref{fig:cartoon_structure}) to consider the non-thermal desorption mechanisms and gas-phase formation of methanol. Inside the disk we use lower values of $X_{\rm ice} + X_{\rm gas} = 10^{-8}$ and a minimum of $10^{-11}$ (i.e. abundance in $D_{\rm out}$ in Fig. \ref{fig:cartoon_structure}). These values are chosen to match the observations of methanol ice  (\citealt{Boogert2015}) and gas (\citealt{Maret2004}; \citealt{Jorgensen2005}) in warm regions of protostars and cold dark clouds (\citealt{Sanhueza2013}; \citealt{Scibelli2021}) for the envelope component, and to match observation and modelling of Class II disks (\citealt{Walsh2014}; \citealt{Booth2021}; \citealt{Gavino2021}) for the disk component. 

The value of 10$^{-6}$ for the envelope is mainly based on the ice abundances found in \cite{Boogert2015}, although there is a factor ${\sim} 3$ spread observed in the methanol ice abundances in protostars (\citealt{Oberg2011}). The gas-phase abundances of methanol in hot cores are uncertain due to the large uncertainty on the warm hydrogen number density. For example \cite{Jorgensen2016} find a value of $1.5 \times 10^{-6}$ for the methanol abundance with respect to total H and argue that this value is an upper limit due to the uncertainty on the column density of total H found from the optically thick continuum. Hence, using a value for the methanol abundance in the inner envelope based on the ice abundances in the outer envelope is the best that can be done.

The photodissociation of methanol by UV radiation is taken into consideration by setting the methanol abundance to zero where the UV optical depth ($\tau_{\rm UV}$) at 1500\,\r{A} (the wavelength at which methanol is effectively photodissociated, \citealt{Heays2017}) is less than 3 (i.e. $A_{\rm v} {\lesssim} 1$ for small grains). This mimics photodissociation due to the UV excess from the accreting protostar or UV produced in fast jet shocks (e.g., HH46, \citealt{vanKempen2009} and in HH211, \citealt{Tabone2021}). We note that this UV excess is not explicitly added to the radiation field since no chemistry model is included: only its penetration depth into the envelope is considered. The UV optical depth is calculated by first calculating the UV radiation field at wavelength of 1500\,\r{A} in RADMC-3D version 2.0\footnote{\url{http://www.ita.uni-heidelberg.de/~dullemond/software/radmc-3d}} with dust ($F_{\rm UV}$) and without dust ($F_{\rm UV,0}$) to be used as a reference. Then, $\tau_{\rm UV}$ is found by $-ln(F_{\rm UV}/F_{\rm UV,0})$ (\citealt{Visser2011}). This procedure results in a thin layer next to the cavity wall where methanol is absent for low mm opacity dust grains (corresponding to high UV opacity, see Fig. \ref{fig:opacities}) and a thicker layer for high mm opacity dust grains (corresponding to low UV opacity). Moreover, at radii $< 1$\,au the methanol abundance is set to zero assuming that methanol gets thermally destroyed very close to the protostar.

\begin{figure*}
    \centering
    \includegraphics[width=0.8\textwidth]{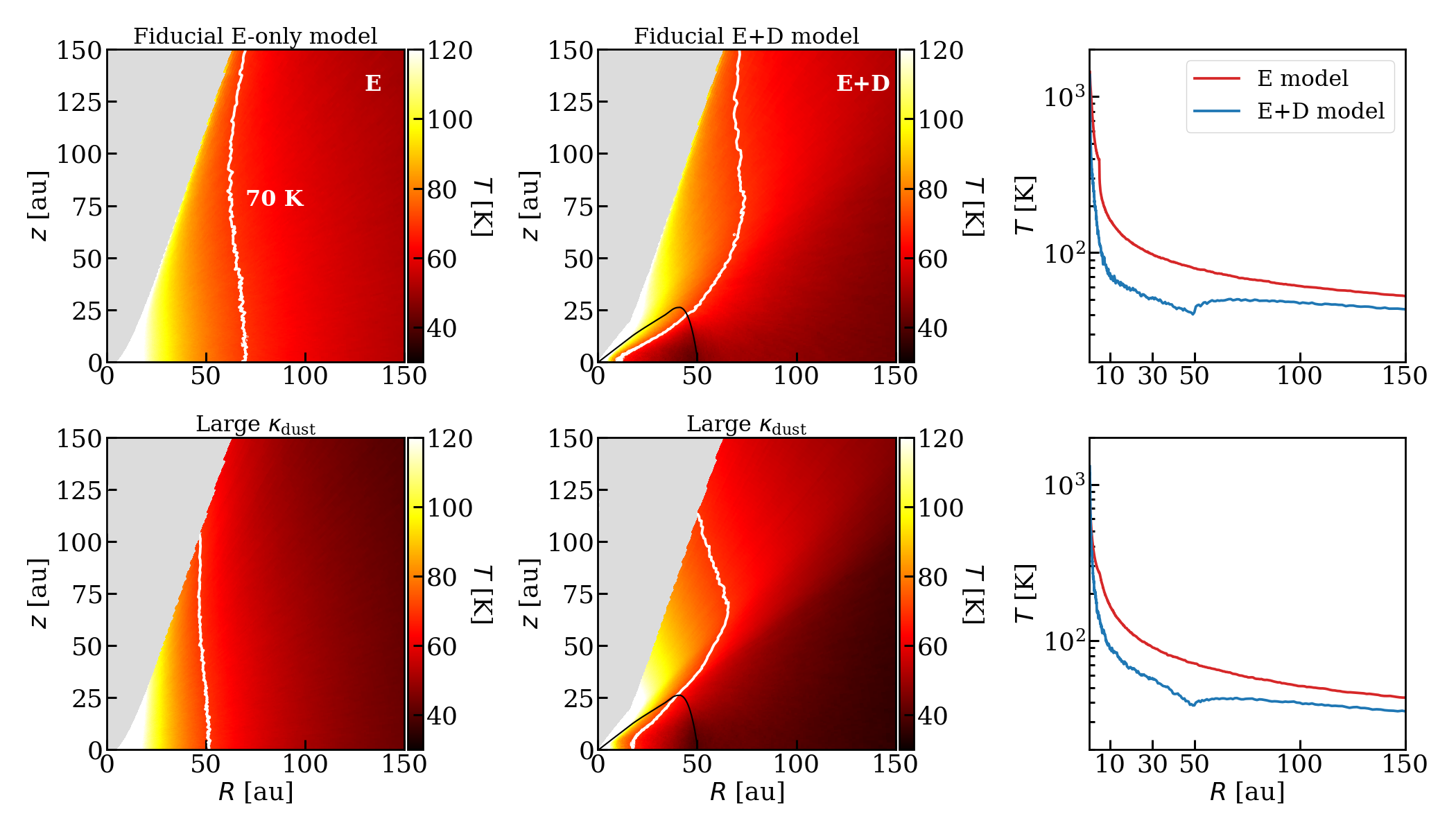}
    \caption{Two-dimensional dust temperatures in the envelope-only models (left column) and in the envelope-plus-disk models (middle column). The white contours show the location where the temperature is 70\,K as an indication of the approximate temperature at which methanol is sublimated at the densities of the models. The right column is the comparison of the mid-plane temperature in the envelope-plus-disk models and the envelope-only models. The first row presents the fiducial model and the second row presents the fiducial model with high mm opacity dust grains. In all figures in this work, $\kappa_{\rm dust}$ refers to opacity of the dust grains at millimetre wavelengths. On average the temperatures are lower when a disk is present and when grains have high mm opacity.} 
    \label{fig:temp}
\end{figure*}

\subsection{Temperature and line emission calculations}

Accurate calculation of the thermal structure of the system is crucial as the temperature sets the sublimation region of methanol and the excitation conditions. The dust and the gas are assumed to be thermally coupled given that the studied regions have high densities. The dust temperature is calculated using Monte Carlo dust continuum radiative transfer by propagation of photons from the central star. We use the code RADMC-3D version 2.0 to do this calculation. The temperature structure and dust opacity depend on the optical properties of the dust. We use two dust distributions one at a time for a model. The two dust distributions encompass the range of variation in opacity at the studied wavelengths (${\sim} 1\,\rm mm$) depending on assumptions on grain size and composition (see Fig. \ref{fig:opacities} and \citealt{Ysard2019}). The first consists of small silicate grains (amorphous olivine) with dust size of 0.1\,$\mu \rm m$ and bulk density of 3.7\,$\rm g\,cm^{-3}$ ($\kappa_{1\,\rm mm}\simeq 0.2$\,cm$^2$\,g$^{-1}$). The second includes larger grains with dust distribution $\propto a^{-3.5}$, maximum dust size of 1\,mm, minimum dust size of 50\,\r{A} and a composition that includes water ice and amorphous carbon in addition to amorphous olivine with an average bulk density over all components of 1.4\,$\rm g\,cm^{-3}$ ($\kappa_{1\,\rm mm}\simeq 18$\,cm$^2$\,g$^{-1}$). For the rest of this paper the former dust distribution is referred to as low mm opacity dust and the latter dust distribution is referred to as high mm opacity dust. The symbol $\kappa_{\rm dust}$ in this paper always refers to opacity at mm wavelengths unless otherwise stated.

The spatial grid is logarithmically spaced in the $r$ direction and linearly spaced in the $\theta$ direction for both envelope-only and envelope-plus-disk models. In the calculation of temperature isotropic scattering is switched on in RADMC-3D (see Fig. \ref{fig:opacities} for the albedo of the dust grains used). The number of grid cells and photons used in calculating the temperature were varied to reach convergence in the temperature structure in the models. The final number of grid cells used in the envelope-plus-disk model in the $r$ direction is 300 for radii between 0.4\,au and 0.5\,au and 700 for radii between 0.5\,au and 10$^{4}$\,au whereas 400 cells are used in the $\theta$ direction. For the envelope-only model 1000 grid cells are used in the $r$ direction and 400 in the $\theta$ direction. Both models assume azimuthal symmetry. In both models the number of photons used for temperature calculation is 10$^6$. Multiple tests showed that this number of photons yields accurate temperatures, while keeping the computational time to a minimum.

After the continuum radiative transfer and temperature calculation, line radiative transfer is done by ray tracing from the protostar to the observer using the RADMC-3D version 2.0 `circ image' command. In this calculation we have assumed local thermodynamic equilibrium (LTE) which is reasonable given the densities inside the snowline (see Appendix \ref{app:crit_density} for more details on the critical density needed for LTE conditions). The effects of non-LTE calculation on the models are further investigated in Sect. \ref{sec:non-LTE}. The molecular data needed in ray tracing of A-CH$_3$OH (as opposed to its E symmetry determined by the nuclear spin alignments of the hydrogen atoms in CH$_3$) is taken from the Leiden Atomic and Molecular Database (\citealt{Schoier2005}). For this file, the energy levels, Einstein $A$ coefficients and transition frequencies were taken from the Cologne Database for Molecular Spectroscopy (CDMS) (\citealt{Muller2001, Muller2005}; \citealt{Xu2008}) and the collisional rate coefficients were taken from \cite{Rabli2010}. The methanol line CH$_3$OH 5$_{1,4}$-4$_{1,3}$ with a frequency of 243.916\,GHz ($E_{\rm up} =$ 49.7\,K, $A_{\rm ij} = 6.0 \times 10^{-5}$) is studied in this work. This line is chosen because it is observed as part of the Perseus ALMA Chemistry Survey (PEACHES) towards low-mass protostars (\citealt{Yang2021}) and hence, it is possible to compare our findings with these observations.

\begin{figure*}
    \centering
    \includegraphics[width=0.6\textwidth]{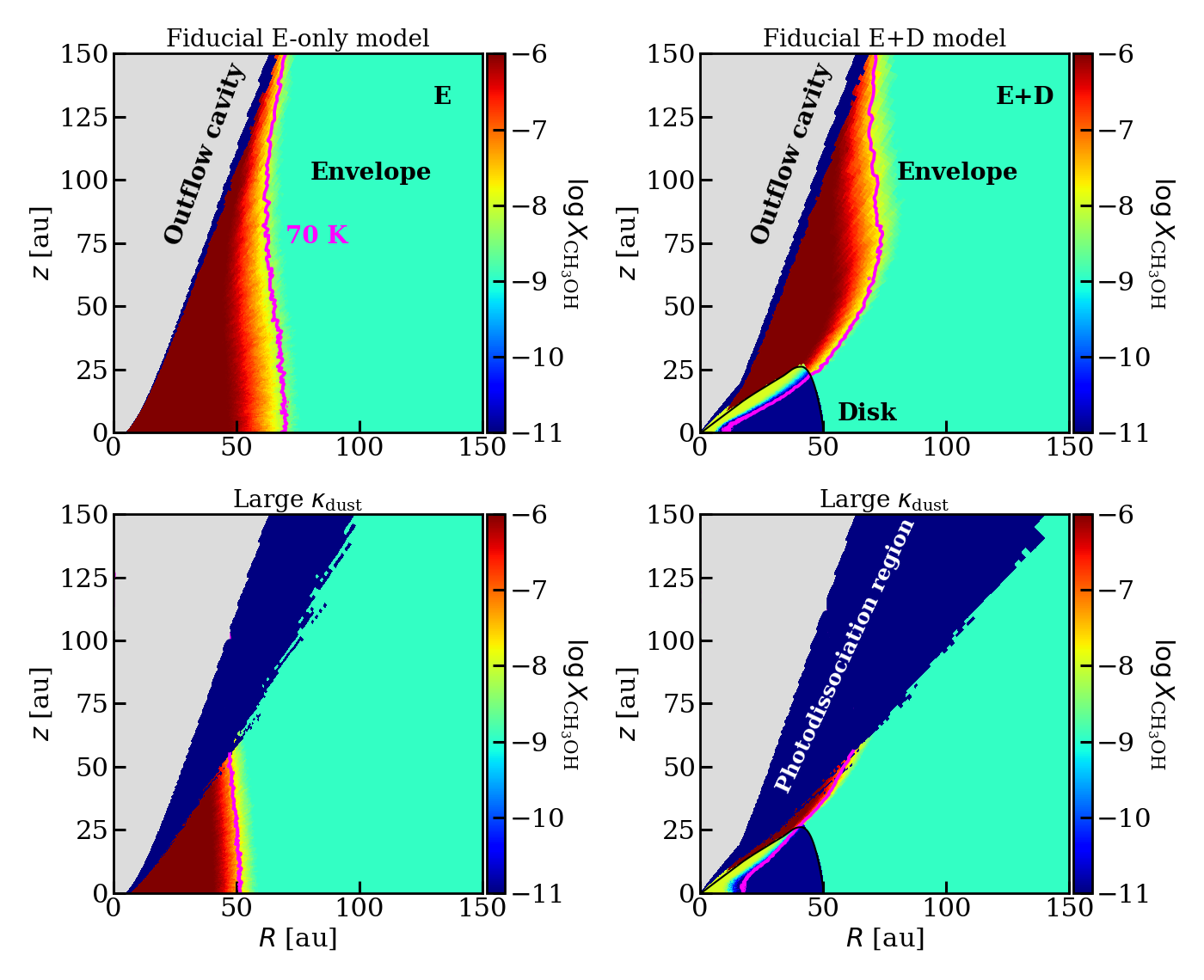}
    \caption{Methanol abundance maps for the envelope-only (left) and envelope-plus-disk (right) models. The magenta contours show where temperature is 70\,K to indicate where the approximate sublimation surfaces of methanol are at the densities in the models. We note the much larger photodissociation region ($\tau_{\rm UV} < 3$) in the envelope for the case when the dust distribution with high mm opacity is used.} 
    \label{fig:meth_abund}
\end{figure*}

To calculate the emission lines, the synthetic images are produced up to velocities of $\pm 15$\,km\,s$^{-1}$ around the centre of the desired methanol line. The velocity bin (i.e. spectral resolution) used in this work is 0.2\,km\,s$^{-1}$. A range of viewing angles are considered going from a completely face-on source to a completely edge-on. Moreover, during the ray tracing both gas and dust are included in the models and later the spectral lines are continuum subtracted when measuring the integrated intensities. Isotropic scattering is switched on for the case of high mm opacity grains and is switched off in the case of low mm opacity grains during the ray tracing. The reason for this is that when the grains have low mm opacity (they are small) there is no difference between the emission lines if the scattering is included or not (see Fig. \ref{fig:opacities}) and switching scattering off will decrease the computation time. Again a convergence test is done to indicate the number of photons needed for scattering during the ray tracing for models with high mm opacity dust grains. The number of photons used for scattering during the ray tracing in the final models is 10$^{5}$. To generate the spectral lines we integrate the emission over an area with 2$\arcsec$ diameter (300\,au at assumed source distance of 150\,pc) to mimic the typical spatial resolution of submillimetre surveys where most sources are unresolved. This assumed area introduces an uncertainty for some models when the integrated flux from the 2$\arcsec$ area is compared to the true integrated flux over which warm methanol is emitting. However, this uncertainty is a factor ${\lesssim} 2$ for most of the models considered in this work. This uncertainty gets larger to a factor ${\sim} 3$ underestimate for some extreme cases, e.g., the models with highest luminosity and small envelope mass. Similarly, the amount is overestimated for example for the lowest luminosity cases with large envelope mass. However, these cases do not dominate the sample of models. The area of the 100\,K radius that \cite{Bisschop2007} use to calculate the beam dilution factor for their observations is for most of the models well within the area assumed here (i.e., the 100\,K radius is usually smaller than a 1\arcsec radius; also see \citealt{vantHoff2021} for low-mass sources).

A turbulent velocity of 1\,km\,s$^{-1}$ is assumed in the envelope component of the two models and 0.1\,km\,s$^{-1}$ is assumed in the disk component of the envelope-plus-disk model. We assume no free-fall velocity for simplicity as we are only interested in the integrated line fluxes rather than the small scale kinematical structure of the envelope. However, in the disk component, a Keplerian velocity ($\sqrt{GM_{\star}/R}$) is assumed. These values in both models result in a full width half maximum (FWHM) of ${\sim}2$\,km\,s$^{-1}$ for the emission lines that are similar to what is observed for young protostars (see \citealt{Jorgensen2005}).

\section{Results}
\label{sec:results}
\subsection{Thermal structure and the snow surfaces}
\label{sec:temp_structure}

Figure \ref{fig:temp} presents the temperature structure of the envelope-only model on the left and the temperature structure of the envelope-plus-disk model in the middle column with the comparison of the radial temperature profiles of the two models through the mid-plane in the right column. The top row shows the temperature structure for the fiducial envelope-only and envelope-plus-disk models: $M_{\rm D} = 0.01$\,M$_{\odot}$, $R_{\rm D} = 50\,\rm au$, $M_{\rm E} = 1$\,M$_{\odot}$ and $L_{\star} = 8$\,L$_{\odot}$ with the low mm opacity dust grains. The bottom row corresponds to the same set of models including large dust grains, representative of dust with large opacity in the millimetre (large $\kappa_{\rm dust}$). 

From the first row third column, one can see that the temperature structure of the envelope-only model approximately follows a power-law (with exponent of ${\sim} -0.5$) in the outer envelope where dust is optically thin to the reprocessed far-IR and millimetre wavelengths from other grains. This is consistent with the expected scaling with the optically thin assumption at the Rayleigh-Jeans limit. However, in the inner envelope where dust is optically thick to the reprocessed shorter wavelengths, radiative transfer modelling becomes crucial and the profile gets steeper than the scaling suggests. This is consistent with what \cite{Jorgensen2002} found in their envelope-only models. In the envelope-plus-disk model, at large radii ($\gtrsim 200$\,au) the temperature profile follows a power-law similar to the envelope-only model but at lower temperatures. At small radii where the disk is located the temperature difference between the two models is more significant.

\begin{figure*}
    \centering
    \includegraphics[width=17cm]{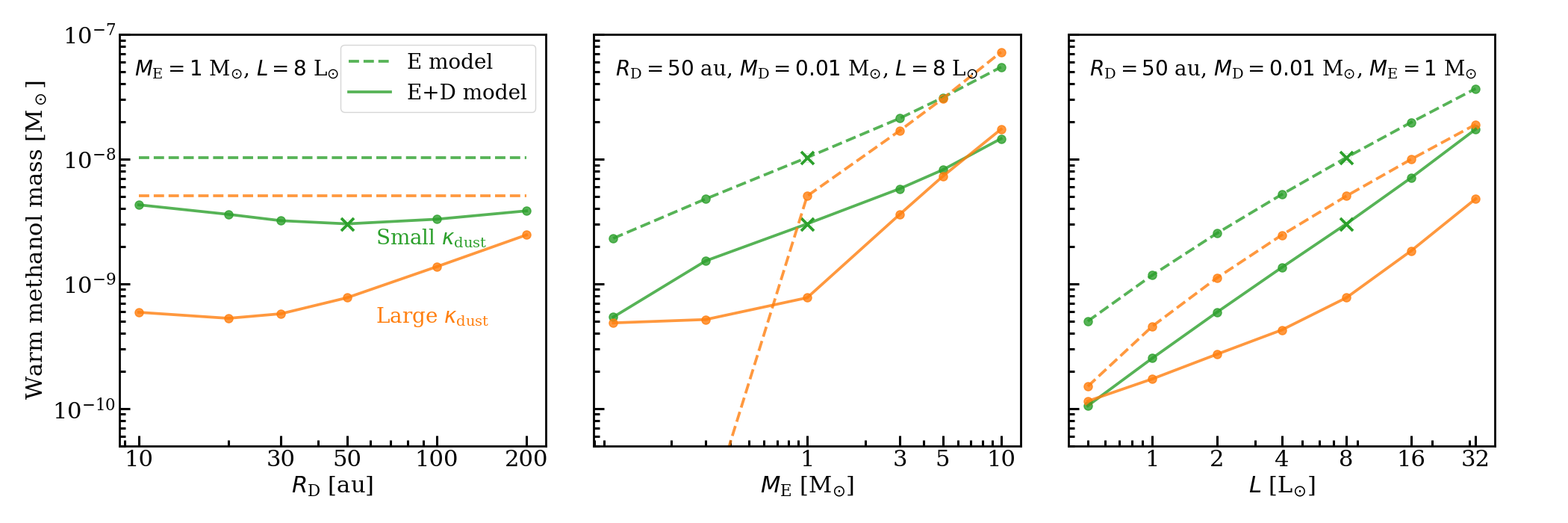}
    \caption{Change in warm methanol mass with disk radius (left panel), envelope mass (middle panel) and luminosity (right panel). The dashed lines show the envelope-only models and the solid lines show the envelope-plus-disk models. The fiducial models are indicated by a cross. Warm methanol mass is computed inside the methanol snow surface where the methanol abundance is larger than 10$^{-9}$ in the envelope and larger than 10$^{-11}$ in the disk (i.e. the methanol mass in $D_{\rm in}$ and $E_{\rm in}$ in Fig. \ref{fig:cartoon_structure}) for models with low mm opacity dust grains (green) and high mm opacity dust grains (orange). In the models where $R_{\rm D}$ is varied, luminosity and mass of the envelope are fixed to 8\,L$_{\odot}$ and 1\,M$_{\odot}$. The reason that the dashed lines are constant in the left panel is that the envelope-only models do not have a disk for which its radius can be altered. Where $M_{\rm E}$ is varied, the luminosity, disk mass and radius are fixed to 8\,L$_{\odot}$, 0.01\,M$_{\odot}$ and 50\,au. Where luminosity is varied, the envelope mass, disk mass and radius are fixed to 1\,M$_{\odot}$, 0.01\,M$_{\odot}$ and 50\,au.} 
    \label{fig:Mass_Rd}
\end{figure*}

In the upper layers of the disk the temperature is the hottest and in the mid-plane it is the coldest. The disk is heated by passive irradiation from the protostar and reprocessing by dust (\citealt{Chiang1997}; \citealt{Dullemond2001}).The effect of disk shadowing is to lower the temperature beyond the outer edge of the disk at radii larger than ${\sim} 50\,\rm au$.

Comparing the two rows, the temperature profiles are similar for the two dust distributions with the methanol snow surface at ${\sim} 70$\,K being slightly closer to the central protostar when the dust has a high mm opacity. This is expected because of two effects. First, dust with high mm opacity is less efficient in absorbing UV and optical light (small dust opacity at UV and optical wavelengths). This especially affects the outflow cavity walls and the disk surface that are directly irradiated by the visible and UV photons from the protostar. Therefore, these regions are colder when the dust has a high mm opacity. Second, once the grains with high mm opacity absorb the radiation at each point they are more efficient at re-emitting it at longer wavelengths compared with the dust that has low mm opacity. The differences in snow surface around the outflow cavity walls are related to the first effect, while those in the disk mid-plane are more related to the second effect. The other parameters of the disk and the envelope such as mass and disk radius also have effects on the temperature structure, these effects are shown in Appendix \ref{sec:grid}.

Figure \ref{fig:meth_abund} shows the gas-phase methanol abundance maps for the same models in Fig. \ref{fig:temp}. This plot shows more clearly where methanol comes off the grains. Especially one should note that most of the warm methanol in the disk is located in the hot upper layers rather than the cold mid-plane, as expected. Moreover, this figure shows the photodissociation region of methanol near the cavity edge, which is larger in spatial extent when the dust grains with high mm opacity are used. These dust grains are more optically thin to the UV radiation.

\subsection{Warm methanol mass}
\label{sec:warm_mass}

In Fig. \ref{fig:temp} the location of the methanol snow surface is shown and in Sect. \ref{sec:temp_structure} its dependence on various model parameters, especially the presence of a disk, are discussed. Due to the differences in the snow surface locations one would expect the warm methanol mass inside the snow surface to vary. The warm methanol mass is a relevant value as it is directly proportional to the methanol emission when the spectral line is optically thin.

Figure \ref{fig:Mass_Rd} presents the warm methanol mass i.e. the amount of methanol inside the snow surface as a function of disk radius (left panel), envelope mass (middle panel) and luminosity (right panel). In the left panel, the disk mass is varied simultaneously with the disk radius to make sure that the value of $M_{\rm D}/R_{\rm D}^{2}$ (an approximation for the disk surface density) stays constant in all models so that one can look only at the effect of disk radius rather than changes in surface density.

The first point to note is that the warm methanol mass increases with envelope mass and luminosity. The relation with respect to luminosity is in the form of a power-law. The slope of this relation in logarithmic space agrees well with the 3/4 exponent in the analytical relation of $M_{\rm{CH_3OH}}\propto L^{3/4}$ (see Appendix B of \citealt{Nazari2021} for derivation; \citealt{vanGelder2022}). The reason for this is that the snow surface moves further from the star when the luminosity increases (Fig. \ref{fig:temp_grid}). In the case of increasing envelope mass but constant luminosity the warm methanol mass increases simply because of the higher density of the envelope so that within the same volume there is more methanol mass. 

The warm methanol mass increases mildly with disk radius when the dust grains have a high mm opacity but stays constant with increasing disk radius when dust grains have a low mm opacity. In the case of low mm opacity grains the warm mass is dominated by the envelope component as shown in Fig. \ref{fig:hot_mass_components}. As the disk radius increases, the snow surface gets closer to the protostar (Fig. \ref{fig:meth_abund_Rs_small}) such that the mass decreases in the envelope component and increases in the disk component so that the total warm mass stays roughly constant. In the case of high mm opacity dust grains the larger photodissociation regions affect the warm mass in the envelope component more than the disk component, thus, the warm mass is dominated by the disk component as seen in Fig. \ref{fig:hot_mass_components_compare}. Therefore, increasing the disk radius will increase the warm mass.

\begin{figure*}
    \centering
    \includegraphics[width=17cm]{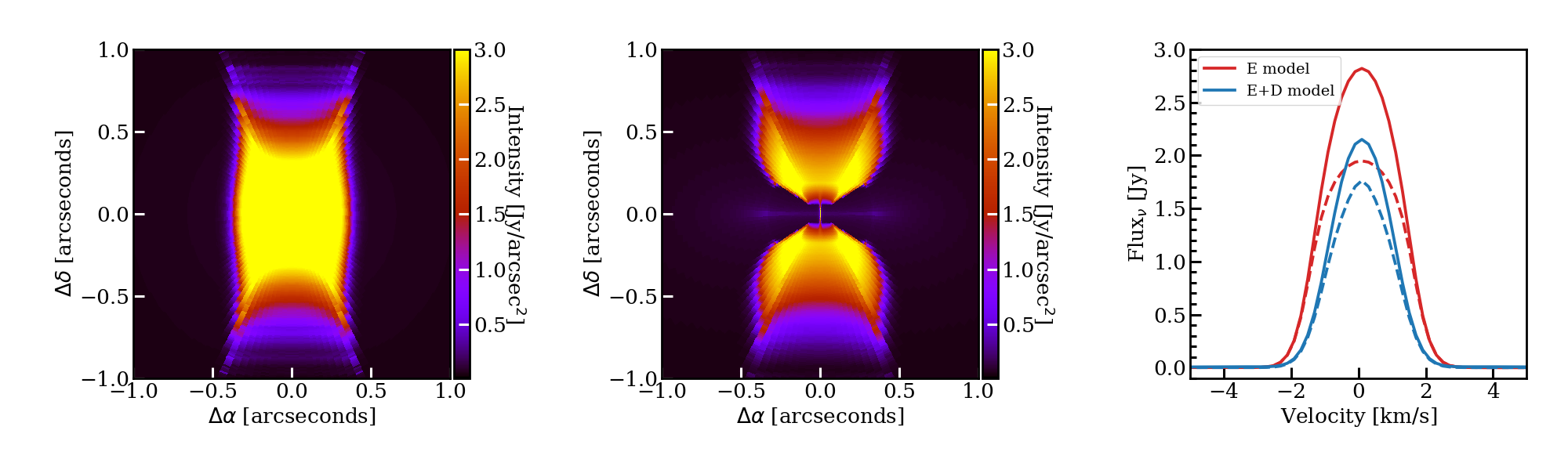}
    \caption{Methanol emission for the fiducial models (low mm opacity dust). The left panel shows the emission from methanol at the peak of the line (0\,km\,s$^{-1}$) for the envelope-only model viewed edge-on, the middle panel shows the same for the envelope-plus-disk model. Given that no free-fall velocity is assumed, the line profile is nearly Gaussian with no infall signatures and thus, the emission at the peak of the line is similar to the integrated intensity map of the line. The right panel is the comparison between the line emission from the envelope-only model and the envelope-plus-disk model seen face-on (dashed lines) and edge-on (solid lines) after integrating over a $2\arcsec$ diameter emitting area (300\,au at the assumed source distance of 150\,pc).} 
    \label{fig:meth_emiss}
\end{figure*}

Focusing on the models with low mm opacity dust grains, the warm methanol mass is always larger in the envelope-only models than the envelope-plus-disk models. This is due to the lower temperatures of the models with disks especially in the disk mid-plane and because of the disk shadowing effect.

In the models with large $\kappa_{\rm dust}$ at mm wavelengths, the warm mass in the envelope-plus-disk models is generally lower than that in the models without disk except for the models with $M_{\rm E}$ of 0.1M$_{\odot}$ and 0.3M$_{\odot}$ (see the large drop in warm methanol mass at low envelope masses in the middle column). In these cases, due to the low envelope masses most of the methanol is photodissociated when the grains have high mm opacity (low UV opacity). However, these envelope masses are on the lower limit/extreme of those observed in Class 0 sources (\citealt{Kristensen2012}).

Comparison between the envelope-only models with small and large $\kappa_{\rm dust}$ at mm shows that the methanol mass is mainly ${\sim} 2$ times lower when the grains have large $\kappa_{\rm dust}$. This is because of two effects. First, the snow surfaces of the models with high mm opacity grains are closer in than those for low mm opacity grains (see Fig. \ref{fig:temp}) but this has a small effect. Second, the photodissociation region of methanol is larger when the mm opacity of the dust is high. However, this drop in warm methanol mass in the envelope-only models between the high and low mm opacity dust grains does not seem to hold when the envelope mass is $> 1$\,M$_{\odot}$ (Fig. \ref{fig:Mass_Rd}, middle panel). This is because the dissociation regions become more similar between the models with high and low mm opacity grains once the envelope mass increases. This is expected as with higher densities it is harder for the UV light to penetrate and at a threshold of ${\sim}3$\,M$_{\odot}$ the photodissociation regions make up a small fraction of the warm envelope (see Fig. \ref{fig:meth_abund_5M}).

Finally, the spread seen in the warm methanol mass for the models shown in Fig. \ref{fig:Mass_Rd} is between ${\sim} 10^{-10}$\,M$_{\odot}$ and ${\sim} 10^{-7}$\,M$_{\odot}$ which is in good agreement with what \cite{vanGelder2022} find from observations of Class 0 objects by ALMA.

\subsection{Methanol emission}
\label{sec:meth_emiss}

\subsubsection{Sample emission lines}

The goal of this work is to compare the methanol emission in the two models with and without a disk. Figure \ref{fig:meth_emiss} shows the line emission from the fiducial models with edge-on view for the line at a frequency of 243.916\,GHz. The left panel shows the emission at 0\,km\,s$^{-1}$ for the envelope-only model observed edge-on, the middle panel shows the same for the envelope-plus-disk model and the right panel shows the respective emission lines seen edge-on (solid lines) and face-on (dashed lines).

The right panel of Fig. \ref{fig:meth_emiss} shows that the peak of line flux for the envelope-only model is ${\sim}$ 1.3 times the flux peak for the envelope-plus-disk model when observed edge-on. Given the similar line widths the integrated line fluxes are also different by a factor ${\sim} 1.3$. This difference can be understood by looking at the amount of warm methanol in the envelope-only and the envelope-plus-disk models in Fig. \ref{fig:Mass_Rd}. For the fiducial models (crosses) in that figure, one can see that the warm methanol mass is ${\sim} 4$ times larger in the envelope-only model than the envelope-plus-disk model and hence one expects a lower methanol flux in the envelope-plus-disk model. The reason that the factor 4 difference in the warm mass has translated to only a factor ${\sim}$ 1.3 difference in the fluxes when observed edge-on, and less when observed face-on, is due to line optical depth effects in the disk and envelope discussed further in Sect. \ref{sec:opacity_intensity}.

\subsubsection{The origin of line emission}
\label{sec:origin_emission}

\begin{figure}
  \resizebox{0.8\hsize}{!}{\includegraphics{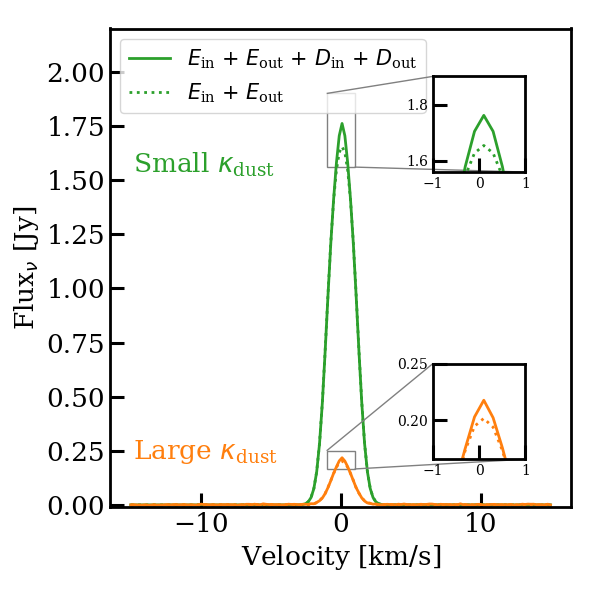}}
  \caption{Comparison of spectral line emission of methanol from the envelope component of the envelope-plus-disk model (dotted lines) and emission from both envelope component and the disk component of the envelope-plus-disk model (solid lines). The line fluxes for the envelope-component of the envelope-plus-disk model (dotted lines) were calculated by setting the abundance of methanol in the disk to zero. Green is for the fiducial model, i.e. low mm opacity dust grains and orange is for the fiducial model with high mm opacity dust grains.}
  \label{fig:emiss_origin}
\end{figure}

Figure \ref{fig:emiss_origin} presents the spectral line emission from the different components of the fiducial envelope-plus-disk model and the same model with high mm opacity dust grains viewed face-on (see Fig. \ref{fig:cartoon_structure} for the components). The emission is dominated by the envelope in both cases of low and high mm opacity dust grains. In the case of low mm opacity dust grains this is due to the small amount of warm methanol in the disk component ($D_{\rm in}$) compared to the envelope component ($E_{\rm in}$). The warm methanol mass from the various components of the fiducial envelope-plus-disk model are presented in Fig. \ref{fig:hot_mass_components}. However, when the grains have high mm opacity the warm methanol mass is dominated by the disk component (see Fig. \ref{fig:hot_mass_components_compare}) but still the emission is dominated by the envelope component. This is because of the continuum over-subtraction effect (explained in Sect. \ref{sec:opacity_intensity}) where emission of methanol from the disk component is `hidden' by the optically thick dust in the disk.  

Moreover, comparing the emission lines when $\kappa_{\rm dust}$ at mm is small with the case where $\kappa_{\rm dust}$ at mm is large, there is a factor of ${\sim} 7$ drop in the peak of line fluxes. Doing the same comparison for the warm methanol mass between the two models (Fig. \ref{fig:Mass_Rd}) shows that the warm mass is only a factor ${\sim} 3.5$ lower when the dust grains have a high mm opacity. Therefore, some part of the factor of ${\sim} 7$ difference is due to the lower amount of warm methanol which in turn is mainly because of larger photodissociation regions when grains have a high mm opacity. The rest of the factor of ${\sim} 7$ difference comes from the larger optical depth of the dust at millimetre wavelengths which blocks the methanol emission in the envelope and hides a significant amount of the methanol emission from the disk (see Sect. \ref{sec:opacity_intensity}).

\subsubsection{Integrated line fluxes}
\label{sec:int_intens}

\begin{figure*}
    \centering
    \includegraphics[width=17cm]{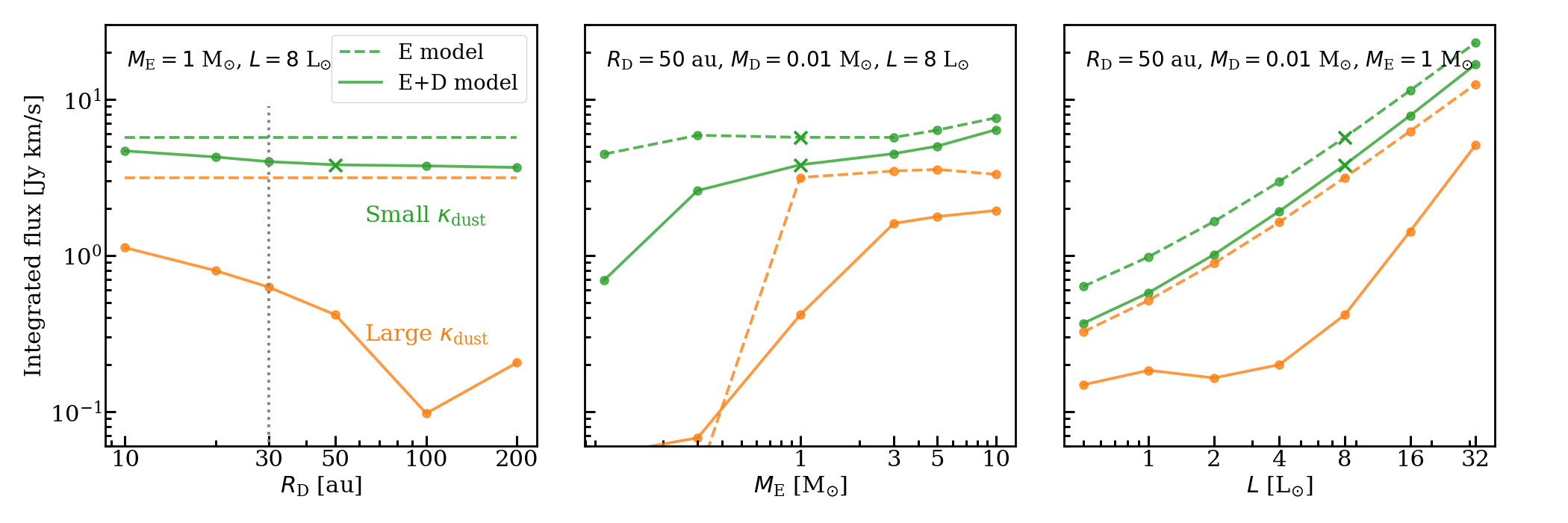}
    \caption{Integrated fluxes of the methanol lines for models with low (green) and high (orange) mm opacity dust grains. The dashed lines show the envelope-only models and the solid lines show the envelope-plus-disk models. The fiducial models are indicated by a cross. The models are the same as those plotted on Fig. \ref{fig:Mass_Rd}. The dotted line indicates the minimum disk radius needed to see a factor ${\sim} 2$ drop in intensity of the envelope-plus-disk model.} 
    \label{fig:int_intensity_Rd}
\end{figure*}

Figure \ref{fig:int_intensity_Rd} presents the integrated fluxes of the continuum subtracted methanol lines for the same models as Fig. \ref{fig:Mass_Rd} as a function of disk radius, envelope mass and luminosity. The effect of viewing angle is examined for the fiducial models. The two extreme cases of viewing angle, face-on and edge-on, show less than a factor 2 difference (Fig. \ref{fig:angle_int} and \ref{fig:angle_int_M5}). Hence, for the rest of this paper only the values for the face-on view are presented.

The trend seen in Fig. \ref{fig:Mass_Rd} as a function of envelope mass and luminosity is reflected in the integrated fluxes. However, the middle panel of Fig. \ref{fig:int_intensity_Rd} shows that the slope of the increase in methanol emission with the envelope mass decreases for envelope masses ${\gtrsim}3$\,M$_{\odot}$. This is due to the line becoming more optically thick (Fig. \ref{fig:tau_line_M3}). It can also be noticed that the line fluxes increase with $L$. While we find that for the fiducial model the emission is always optically thick for the various luminosities (Fig. \ref{fig:tau_line_L}), increasing the luminosity leads to an increase in the area of the emitting region due to the snowline moving outwards, leading to an increase in the integrated flux. 

Moreover, for the low mm opacity grains the integrated fluxes stay constant as the disk radius increases which is the same trend that is seen for the warm methanol masses (Fig. \ref{fig:Mass_Rd}, left panel) but for the high mm opacity grains the fluxes decrease with increasing disk radius which is the opposite of what is seen for the warm methanol masses. This difference can only be explained by the continuum over-subtraction effect discussed in Sect. \ref{sec:opacity_intensity}. As explained in Sect. \ref{sec:warm_mass} a large part of the warm methanol mass comes from the disk component when the grains have a large mm opacity. Hence, if the gas in the disk is optically thick (Fig. \ref{fig:tau_line} and \ref{fig:tau_line_M3}) and the dust in the disk as well (high mm opacity dust grains) the emission from the gas cannot be observed because the line is not brighter than the underlying continuum. 

Comparing the envelope-only models with high and low mm opacity dust grains (the two dashed lines) one can see that there is a drop in integrated fluxes when the grains have a higher mm opacity. In addition to the lower warm methanol mass already described in Sect. \ref{sec:warm_mass}, a radiative transfer effect that lowers even further the emission is dust attenuation in the envelope. This can be understood by the large dust grains being more optically thick at millimetre wavelengths, thus, they block the emission in the envelope.

For both low and high mm opacity grains the envelope-only models have a larger integrated flux than the envelope-plus-disk models. This difference is more prominent when the grains have a high mm opacity. The only exception are models in which the envelope masses are 0.1 and 0.3\,M$_{\odot}$ and the grains have a high mm opacity. These trends are similar to what was seen in Fig. \ref{fig:Mass_Rd}.

From the left panel of Fig. \ref{fig:int_intensity_Rd}, in both low and high mm opacity dust models the integrated line fluxes in the envelope-plus-disk models drop by a factor larger than ${\sim} 1.5-3$ compared with the envelope-only models when the disk radius is at least 30\,au. This disk radius would be smaller if the luminosity is lower and it would be larger if the luminosity is higher. This is quite interesting as this could indicate a minimum disk size for sources with observed weak methanol emission compared with sources that have stronger methanol emission. Recall that $R_{\rm D}$ is not the same as the observed gas or dust disk radii. Moreover, once $\kappa_{\rm dust}$ at mm is large the drop in integrated fluxes between the envelope-only and envelope-plus-disk models increases to more than an order of magnitude for disks with radii larger than 50\,au (Fig. \ref{fig:int_intensity_Rd} left panel). The reason that this drop for large disk sizes is much larger for high mm opacity dust grains than low mm opacity dust grains is the continuum over-subtraction in the optically thick dust disk discussed further in Sect. \ref{sec:opacity_intensity} (Fig. \ref{fig:continuum_oversub_200au}).

\subsection{Caveats}

The abundances of methanol in the inner/outer envelope are parameterised with values matching the observations. However, the observed values vary by a factor ${\sim}$ three for low-mass young stellar objects ($3.3^{7.9}_{2.7} \times 10^{-6}$, \citealt{Boogert2015}; also see \citealt{Bottinelli2010}; \citealt{Oberg2011}). Moreover, \cite{Drozdovskaya2014} find that the abundance of methanol gas and ice in the disk can change during the infall process. The values used in the disk are also parameterised and are on the lower side of those inspired by observations (\citealt{Booth2021}) and chemical modelling of Class II disks (\citealt{Walsh2014}). However, if these values are increased by an order of magnitude the emission lines in the disk become more optically thick (Fig. \ref{fig:tau_line_high_abund}) but they still show values lower than the envelope-only models as explained in Sect. \ref{sec:opacity_intensity}. For example, if the methanol abundance in the disk component of the fiducial envelope-plus-disk model is increased by an order of magnitude the integrated line flux would be only a factor of ${\sim} 1.04$ times higher than that for the fiducial envelope-plus-disk model.

There are a number of assumptions in the physical structures of the models. Here are the most important of them with explanation of their effects on the results of this work:

\begin{itemize}
    \item Gas density structure: The gas density structure for the envelope-only model (Eq. \ref{eq:density_env}) and the envelope component of the envelope-plus-disk model (Eq. \ref{eq:dens}) are not exactly the same. More precisely, at the inner radii just above where the disk is located in the envelope-plus-disk model, the envelope-only model is more dense than the envelope-plus-disk model. Therefore, part of the reason that the envelope-only models show more warm methanol mass is the fact that these models have higher density in the inner envelope. However, the difference between the densities in the inner envelope is at most a factor ${\sim} 5$ at radii of ${\sim} 20$\,au and for most radii below a factor 5 (see Fig. \ref{fig:dens_flattened}). Hence, this cannot be the main reason why the envelope-only models have more than an order of magnitude stronger emission than the envelope-plus-disk models.
    
    \item Shape of the outflow cavity: It has been assumed in the models that the outflow cavity has the same shape in both the envelope-only and envelope-plus-disk models and is characterised by Eq. \eqref{eq:cos_theta0}. Moreover, the outflow cavity opening angle and extent is the same in all the models. However, shape of the cavity at small scales and the mechanism that opens it is still debated and not fully understood. Furthermore, larger/smaller outflow cavities can affect the values reported in this work by enhancing/quenching the UV penetration depth (\citealt{Drozdovskaya2015}). Nevertheless, the main conclusion would stay the same that the envelope-plus-disk models show lower methanol emission due to the lower temperatures.

    \item Disk scale height: The disk scale height is parameterised and hence it is not self consistent with the disk's thermal structure. To examine the effect of disk scale height on the integrated line fluxes, the scale height is calculated using the mid-plane temperature of the fiducial model and the fiducial model is run with the new consistent scale height. This results in an $H/R$ of ${\sim}0.06$ at 1\,au compared with $0.2$ used for the fiducial model. It is important to note that in reality the mid-plane temperature refers to that of dust, whereas flaring depends on the gas temperatures which decouples from those of dust in the surface layers. One should calculate the gas temperature self consistently to obtain realistic structures. However, this calculation is computationally expensive and beyond the scope of this work thus, the mid-plane temperature is used. When the mid-plane temperature is used to find the scale height the integrated flux is only a factor ${\sim 1.3}$ larger than what is assumed for the fiducial model. Therefore, its effect should not change the final conclusions of this work.
    
    \item Disk extent: The disk is assumed to have a hard edge. The material beyond the disk edge is cold and thus does not contribute much to methanol emission. Hence, adding a disk without a hard edge (e.g. an exponentially decaying density structure) will only change the amount of cold methanol in the models, and not affect the methanol fluxes significantly. Furthermore, the effect of having a more extended disk on methanol emission is to decrease the emission in the envelope-plus-disk models because there will be a larger area in which the methanol abundance is $10^{-11}$ rather than $10^{-9}$. Therefore, the difference between the envelope-only model and the envelope-plus-disk models will get larger. 

    \item Free-fall velocity: No free-fall velocity is assumed in the models. Fiducial envelope-only and envelope-plus-disk models have been run after adding free-fall velocity to the envelope component (the models are not shown here). The integrated flux for the envelope-only case is ${\sim} 2$ times larger and for the envelope-plus-disk case is a factor of ${\sim} 1.3$ larger when free-fall velocity is added. This shows that adding free-fall velocity will only increase the difference between the envelope-only and the envelope-plus-disk models.  
\end{itemize}

\section{Discussion}
\label{sec:discussion}
\subsection{Opacity effects}
\label{sec:opacity_intensity}

In the previous section it is shown that the methanol emission from the envelope-plus-disk models is always lower than the envelope-only models. This is mainly due to the smaller warm methanol mass in the envelope-plus-disk models as the temperatures are lower and hence the snow surfaces are closer to the central protostar.

The line optical depth effects can also become important. The fiducial model of envelope-plus-disk for the considered spectral line has optically thick methanol ($\tau_{\rm CH_{3}OH} {\gtrsim} 0.5$) in the inner ${\sim} 50$\,au and the fiducial envelope-only model has optically thick methanol between radii of ${\sim} 5$\,au and ${\sim} 60$\,au (Fig. \ref{fig:tau_line}). Hence, in these regions the line flux is proportional to the emitting area which is smaller in the envelope-plus-disk model as the snow surface lies closer to the protostar due to the presence of a disk and disk shadowing (Fig. \ref{fig:temp}). Therefore, a factor ${\sim} 3.5$ drop in warm methanol mass has translated into a factor ${\sim} 1.5$ drop in the integrated fluxes between the fiducial models of envelope-only and envelope-plus-disk. It should also be noted that the line is more optically thick in the disk than the envelope for the fiducial envelope-plus-disk model, especially in the inner 10\,au. The line optical depth effects become especially important if larger methanol abundances are assumed in the disk and the envelope. If the methanol abundance in the disk is higher by an order of magnitude for the fiducial envelope-plus-disk model, the envelope-only model would still have stronger methanol emission than the envelope-plus-disk model due to a smaller emitting area (see Fig. \ref{fig:tau_line_high_abund}).   

As explained in Sect. \ref{sec:int_intens} the drop in integrated fluxes are much larger between the two models when the dust grains have a high mm opacity. The temperature difference between the models with low and high mm opacity dust grains is small (Fig. \ref{fig:temp}). For example the warm methanol mass is the same for the case where $M_{\rm E} = 5$\,M$_{\odot}$ between the envelope-only models with low and high mm opacity dust grains (Fig. \ref{fig:Mass_Rd}) while there is a factor ${\sim} 2$ difference between the integrated line fluxes of these two models . This points towards the effect of dust opacity rather than the temperature structure. The dust opacity has two types of effects: dust attenuation in the envelope and continuum over-subtraction in the disk.

\begin{figure}
  \resizebox{\hsize}{!}{\includegraphics{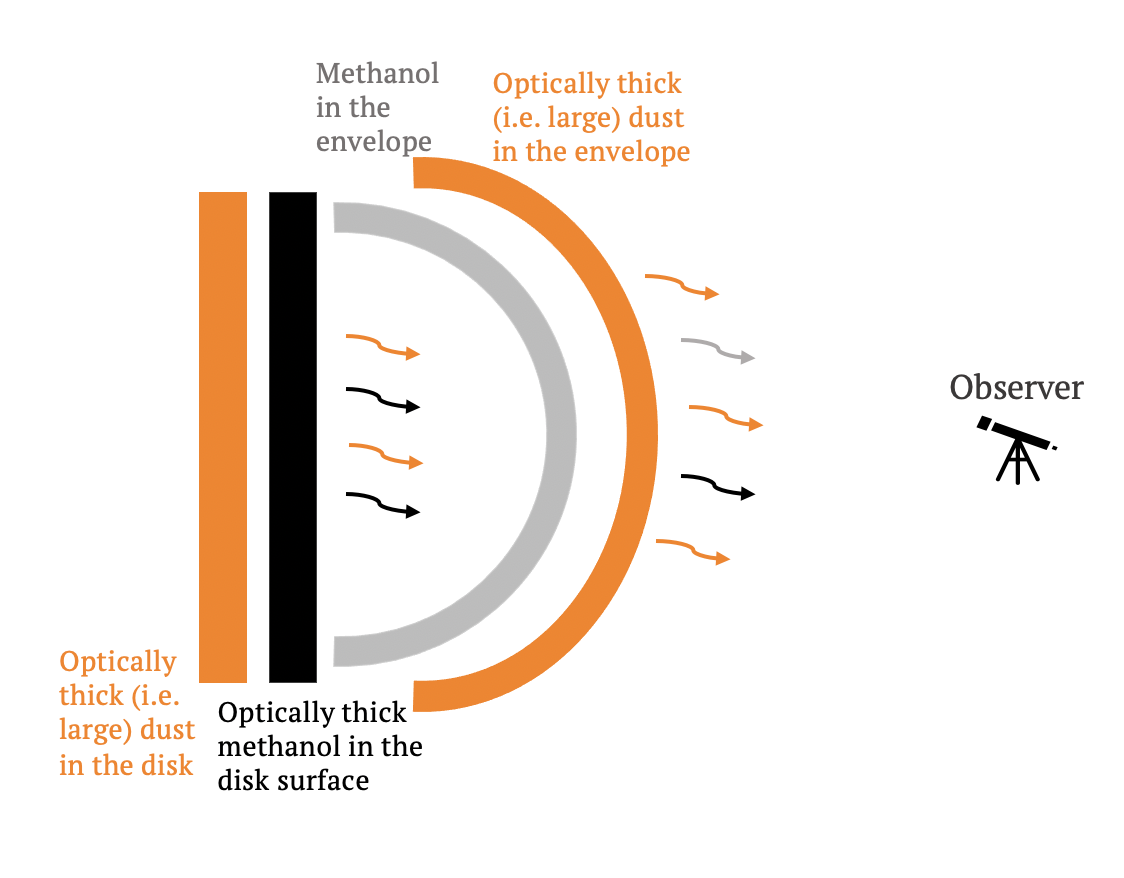}}
  \caption{A sketch of the emission from both methanol and dust in the disk and envelope of the envelope-plus-disk model to emphasise the effects of dust optical depth in the disk and the envelope. In the envelope the dust can block the methanol emission whereas in the disk, although methanol is in front of the dust, the dust optical depth causes an error due to continuum over-subtraction.}
  \label{fig:optical_sketch}
\end{figure}

Dust attenuation in the envelope is happening on scales of ${\sim}100$\,au in the envelope-only model. The dust with high mm opacity becomes marginally optically thick (${\gtrsim} 0.1$, see Fig. \ref{fig:tau_fiducial_largedust}) for envelope masses ${\gtrsim} 1$\,M$_{\odot}$. This phenomenon in addition to the larger photodissociation regions are the two main reasons for the drop in integrated fluxes seen between the two dashed lines in Fig. \ref{fig:int_intensity_Rd} (all panels). A cartoon of this effect can be seen in Fig. \ref{fig:optical_sketch}, especially looking at the envelope components of the model (the two curved lines). This figure shows that the optically thick dust can be between the methanol emission in the envelope and the observer and hence is blocking the methanol emission.  

The continuum over-subtraction effect only happens in the envelope-plus-disk model. This effect has been discussed in the literature (\citealt{Boehler2017}; \citealt{Weaver2018}; \citealt{Rosotti2021}) and states that continuum subtraction can cause underestimation of the true line intensity if both the dust and the gas in front of the dust in the disk are optically thick and the line has a similar brightness to or lower than the underlying continuum. The total intensity at frequency $\nu$ is the sum of the disk component intensity ($I_{\nu,D}$) and the envelope component intensity ($I_{\nu,E}$) and is given by

\begin{align}
\label{eq:I_nu}
\begin{split}
    I_{\nu} = & ~   I_{\nu,D} + I_{\nu,E} = \\ & (1 - e^{-\tau_{\rm d,D}}) e^{-\tau_{\rm g,D}} e^{-\tau_{\rm g,E}} e^{-\tau_{\rm d,E}} B_{\nu} (T_{\rm d,D})\\& + (1 - e^{-\tau_{\rm g,D}}) e^{-\tau_{\rm g,E}} e^{-\tau_{\rm d,E}} B_{\nu} (T_{\rm g,D}) \\& + (1 - e^{-\tau_{\rm g,E}}) e^{-\tau_{\rm d,E}} B_{\nu} (T_{\rm g,E})\\& + (1 - e^{-\tau_{\rm d,E}}) B_{\nu} (T_{\rm d,E}),
    \end{split}
\end{align}

\noindent where $\tau$ is the optical depth, $B_{\nu}$ is the Planck function and $T$ is the temperature. Subscripts d and g denote dust and gas, where subscripts E and D denote the envelope and disk components respectively. Here four layers of dust and gas are assumed as shown schematically in Fig. \ref{fig:optical_sketch}.

This equation assumes that both the gas and dust in the disk and envelope have a finite optical depth. This means that apart from the emission from each layer separately, the optically thick dust in the envelope absorbs some of the dust and gas emission behind it (the $e^{-\tau_{\rm d,E}}$ terms), the optically thick gas in the envelope absorbs some of the dust and gas emission behind it (the $e^{-\tau_{\rm g,E}}$ terms) and the gas in the disk absorbs the dust emission behind it (the $e^{-\tau_{\rm g,D}}$ term). If the line and dust emission are optically thin, after the continuum subtraction (i.e. subtracting the first and fourth terms), $I_{\nu}$ will be the true line intensity. However, if the dust and line emission are optically thick in the disk, the gas in the disk will absorb part of the dust emission ($e^{-\tau_{\rm g,D}}$ in the first term) and does not emerge over the continuum, therefore, subtracting the dust intensity (first term) will eliminate some (or all) of the emission from the line in the disk.

\begin{figure}
  \resizebox{\hsize}{!}{\includegraphics{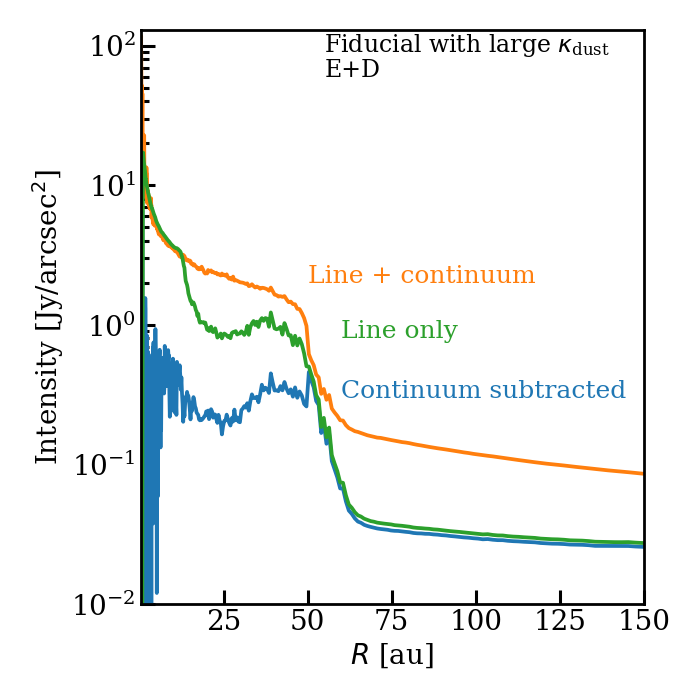}}
  \caption{Effect of continuum over-subtraction in the fiducial envelope-plus-disk model with large $\kappa_{\rm dust}$ at mm. Orange shows a radial cut through methanol emission image including dust continuum at the line's peak (0 km s$^{-1}$). Green shows the same where there is no dust in the model (true line intensity). Blue shows the same as orange where the continuum is subtracted. There is a factor ${\sim} 4$ to up to an order of magnitude difference between the continuum subtracted intensity and the true line intensity due to the continuum over-subtraction effect.}
  \label{fig:continuum_oversub}
\end{figure}

This effect of the continuum over-subtraction can be quantified in Fig. \ref{fig:continuum_oversub}. This figure shows a cut through the image of the peak of the spectral line observed face-on for the envelope-plus-disk fiducial model with optically thick dust. Orange shows the cut through the image without continuum subtraction, blue shows the same after continuum subtraction and green shows the same with no dust included in calculation of the image. At radii below ${\sim} 25$\,au there is about an order of magnitude difference between the true line intensity (green) and the continuum subtracted intensity (blue). This difference is smaller for radii between 25\,au and 50\,au but it is still a factor ${\sim} 4$. However, there is no difference between line only and continuum subtracted intensity where the disk stops (after 50\,au). This figure shows that the continuum over-subtraction effect can lead to a factor 4-10 error in measuring the true line intensity. 

To conclude, the dust can have three effects. It either blocks the emission in the envelope or it hides the methanol emission and introduces an error when the emission is continuum subtracted. Moreover, the photodissociation regions are larger when high mm opacity dust grains are used as they are more optically thin to UV compared with low mm opacity dust grains.

\subsection{Non-LTE effects}
\label{sec:non-LTE}

Throughout this paper, the emission of methanol is computed under the LTE assumption. Critical densities of the low-lying methanol levels ($E_{\text{up}} {\lesssim} 100~$K) are typically found to be of the order of $n_{\rm crit} \simeq 10^{6}$~cm$^{-3}$ (Fig. \ref{fig:crit_dens}). One could then expect that non-LTE effects might play a role in the low density part of the envelope. In order to test the validity of our LTE approximation, we ran a series of non-LTE models using the Ratran modelling code \citep{2000A&A...362..697H}, taking into account collision with H$_2$ \citep{2010MNRAS.406...95R} and radiative pumping. For the purpose of this work and in order to save computational time, the envelope is treated as a 1D spherically symmetric structure. The density follows the prescriptions detailed in Sect. \ref{sec:env_model} except that the outflow cavity is discarded. This provides a conservative estimation of the non-LTE effects as in the presence of a disk, methanol emission is confined to even denser regions, where LTE should be more appropriate. The thermal structure has been computed with RADMC-3D and the abundance of methanol is set as described in Sect. \ref{sec:methods}. For conciseness, only models with high mm opacity dust grains and $L=8$\,L$_{\odot}$ have been computed.

\begin{figure}
  \resizebox{\hsize}{!}{\includegraphics{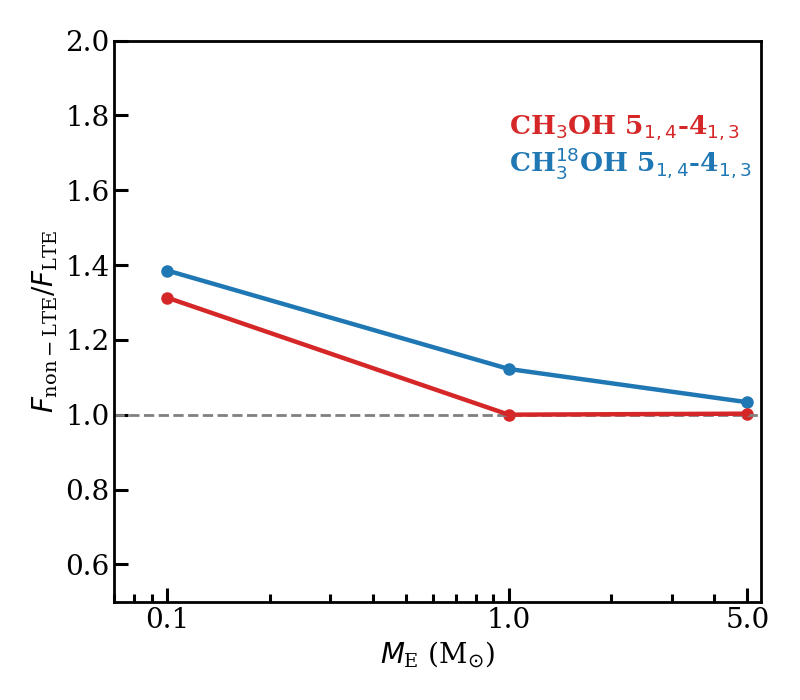}}
  \caption{Impact of non-LTE effect on the emission of the $5_{1,4}$-$4_{1,3}$ transition of methanol ($E_{\rm up} =$ 49.7\,K, $A_{\rm ij} = 6.0 \times 10^{-5}$) emitted from the inner 2$\arcsec$ of the envelope. The y-axis indicates the ratio between the integrated line flux computed with Ratran under non-LTE assumptions and LTE assumptions. The non-LTE computation is done with a consistent calculation of the population levels of methanol, including collision with H$_2$ and radiative pumping. The line flux from CH$_3^{18}$OH is computed assuming the same collisional rate coefficients and an isotopologue ratio of $^{16}$O/$^{18}$O$= 560$.}
  \label{fig:non-LTE}
\end{figure} 

Figure \ref{fig:non-LTE} shows the ratio of integrated line fluxes between two models, one computed with non-LTE assumptions and the other with LTE assumptions as a function of envelope mass. Red shows this ratio for CH$_3$OH and blue for CH$_{3}^{18}$OH to consider the non-LTE effects on an optically thin line. It demonstrates that non-LTE effects do not impact significantly the intensity of the CH$_3$OH 5$_{1,4}$-4$_{1,3}$ line (red) down to an envelope mass of $0.1 M_{\odot}$. The emission here is dominated by the region interior to the sublimation surface that has densities above $\sim 10^{7}~$cm$^{-3}$. We note that \cite{Jorgensen2005} and \cite{Maret2005} performed full non-LTE calculations to model single dish CH$_3$OH emission, which was necessary in their case because their observed emission originates both from the hot core and the larger scale cold, lower density envelope.

Other than the high densities of the inner regions, the large opacity of the strong methanol line considered here also contributes to quench the radiative de-excitation of the level, leading to LTE populations at densities even below the critical densities of the levels ("photon trapping"), which can help explaining the small difference between LTE and non-LTE models in Fig. \ref{fig:non-LTE}. Still, for the CH$_3^{18}$OH 5$_{1,4}$-4$_{1,3}$ line (blue), that is not affected by photon trapping (optically thin lines), the LTE assumption is largely valid since non-LTE calculations are only ${\sim}20\%$ different. Interestingly, the non-LTE effect tends to enhance the line emission for this specific line. We also note that for more excited lines ($E_{\rm up} \gtrsim 300~$K), non-LTE effects start to be important. This was confirmed by checking the population of an excited level which showed substantial deviation from LTE.

\subsection{Comparison with observations}

\begin{figure}
  \resizebox{0.9\hsize}{!}{\includegraphics{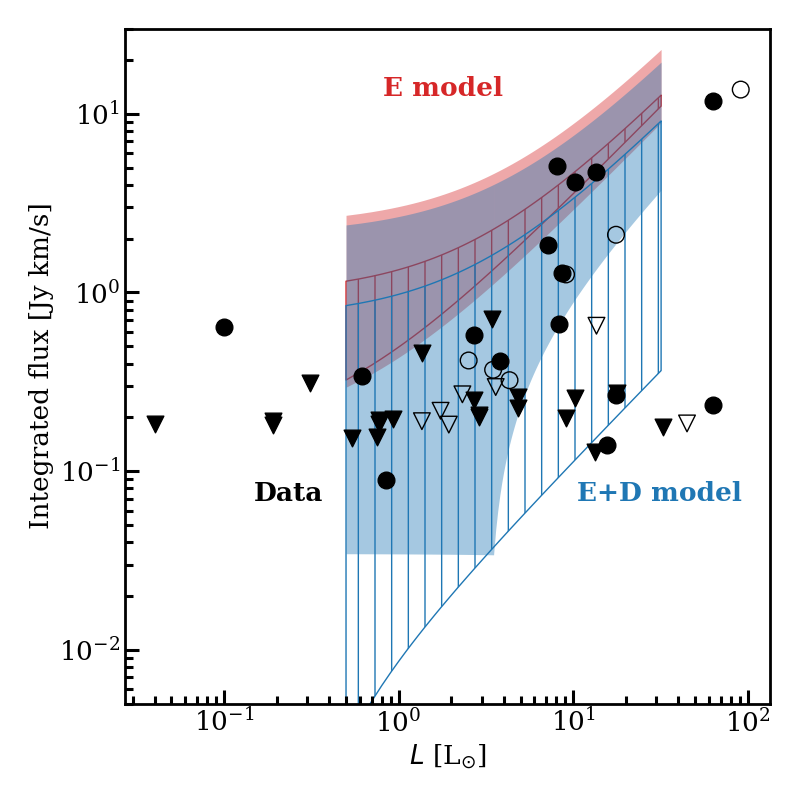}}
  \caption{Comparison of the integrated intensities found from the models with observations (black circles) normalised to a source distance of 150\,pc. Red represents the regions with envelope-only models and blue shows the regions with envelope-plus-disk models. The smooth regions indicate the models with low mm opacity dust grains and striped regions show the models with high mm opacity dust grains. The envelope-only models with high mm opacity dust grains and envelope masses of 0.1\,M$_{\odot}$ and 0.3\,M$_{\odot}$ are not included here as most of the envelope is photodissociated (Sect. \ref{sec:warm_mass}). Triangles show the upper limits found from observations and the hollow points are the sources found to have a disk. To explain most of the data both a disk and optically thick dust are needed.}
  \label{fig:comp_obs}
\end{figure}

We now address the main question of this paper: Whether source structure can explain the lack of methanol emission observed for some sources. Figure \ref{fig:comp_obs} shows integrated fluxes of the CH$_3$OH $5_{1,4}$-$4_{1,3}$ transition at 243.916\,GHz from the PEACHES survey (\citealt{Yang2021}; \citealt{vanGelder2022}) in black with the envelope-only models in red and envelope-plus-disk models in blue. The smooth coloured regions show where models with low mm opacity dust fall and the striped regions show the same for models with large mm opacity dust.

One can see that the envelope-only models cannot explain the bulk of the observations, for the adopted methanol abundances. They can only explain the observations with high methanol fluxes but cannot explain the sources with low methanol emission even when the dust is highly optically thick. This is important because it shows that thinking about protostars as envelope-only objects is not necessarily correct. The CH$_3$OH abundance would have to be up to two orders of magnitude lower to explain the range. This is inconsistent with ice observations of dense clouds assuming that all the ices sublimate inside the snow surface with no subsequent chemistry taking place.

The envelope-plus-disk models, however, can explain most of the observations as they have lower integrated line fluxes. There is a group of sources with very low methanol emission that only models with disk and high mm opacity dust grains (high dust optical depths) can reproduce. Therefore, to understand the observations one needs to take into account both the disk and dust optical depth effects, i.e. dust grains with large $\kappa_{\rm dust}$ at mm.  

Based on the results from this work one can conclude that most of the sources without strong CH$_3$OH emission observed as part of the PEACHES survey (\citealt{Yang2021}) are prime targets to find disks in low-mass protostars. To test this hypothesis one can look at cold species such as DCO$^{+}$ (as done by \citealt{Murillo2018}) in the protostars observed by the PEACHES survey. Moreover, observations of these sources at longer wavelengths (as done by \citealt{DeSimone2020} for NGC 1333 IRAS 4A1) can help distinguish between non-existence of complex organics in a source and the dust optical depth effects in the disk and the envelope. Furthermore, very high angular resolution observations (${\sim} 30$\,au) can help resolve very small disks in the continuum.

The hollow markers on Fig. \ref{fig:comp_obs} indicate the sources that are confirmed to have a disk from the VLA Nascent
Disk and Multiplicity (VANDAM) survey (\citealt{Segura-Cox2018}). It is reassuring that all cannot be explained by the envelope-only models. There are other examples in the literature where sources with disks show no or less amount of complex organic molecules. For example, \cite{Tychoniec2021} observe a large resolved dust structure in the Class 0 source, Serpens SMM3, perpendicular to the outflow which could potentially be a disk and find no strong methanol or complex species towards this source despite the high luminosity of the source ($L_{\rm bol} \simeq 28$\,L$_{\odot}$). \cite{ArturdelaVillarmois2019} found no methanol emission towards 12 Class I protostars hosting disks that they observed in the Ophiuchus molecular cloud. Moreover, \cite{vantHoff2020c} do not detect methanol towards two young disks L1527 IRS and IRAS 04302+2247 (also see \citealt{Podio2020}). In addition, \cite{Lee2017,Lee2019} find that the presence of a disk can affect the emission from complex organic molecules in HH212.

Are there alternative explanations for the large spread?  This paper does not model the CH$_3$OH chemistry explicitly but uses parameterised abundances inspired by observations and detailed gas-grain chemistry models. \cite{Notsu2021} study the effect of X-rays on molecular abundances in the inner warm envelope, most notably H$_2$O and CH$_3$OH. Some low-mass Class 0 sources are known to be strong X-ray emitters (e.g., \citealt{Forbrich2006}; \citealt{Grosso2020}), and for
high X-ray luminosities, L$_{\rm X}\gtrsim 10^{30} $\,erg s$^{-1}$, gaseous H$_2$O and CH$_3$OH are efficiently destroyed within their snow lines. This could result in low CH$_3$OH line intensities for some of our sources. X-ray destruction can be distinguished from the scenarios discussed here through
observations of HCO$^+$ and its optically thin H$^{13}$CO$^+$ isotopologue (see also \citealt{vantHoff2021}): if X-ray destruction dominates, HCO$^+$ emission would be centrally peaked on source, having a high abundance even within the water snow line since its main destroyer (water) is absent. \cite{Notsu2021}, see their section 4.6, identify two sources in the Perseus Class 0 sample with weak CH$_3$OH emission and centrally peaked HCO$^+$ for which this could be the case. Additional deep surveys of both CH3OH and H$^{13}$CO$^+$ of this sample is needed to assess how common this explanation is.

\section{Conclusions}
\label{sec:conslusion}
This work investigates whether the presence of disks can explain the lack of methanol emission from some low-mass protostellar systems. Two models are considered here: a spherically symmetric envelope-only model with an outflow cavity and a flattened envelope model with an embedded disk and an outflow cavity. Radiative transfer calculations are performed for these two models using the radiative transfer code RADMC-3D to calculate the thermal structure and next, methanol emission assuming parameterised abundances and LTE excitation. The main conclusions of this work can be summarised as follows:

\begin{itemize}
    \item The envelope-plus-disk models always show lower methanol emission than the envelope-only models. This is mainly due to disk shadowing and lower temperatures in the disk mid-plane. These two effects result in smaller sublimation regions and hence, a lower warm methanol mass in the envelope-plus-disk model. 
    \item The methanol emission from the envelope-plus-disk model is weaker than the envelope-only models by a factor ${\sim} 2$ if the disk is at least 30\,au in size (for $L=8$\,L$_{\odot}$).
    \item The drop in intensities between the two models gets larger than one order of magnitude when dust grains have a high mm opacity and the disk is at least 50\,au in size (for $L=8$\,L$_{\odot}$). This is because of the continuum over-subtraction in the disk. 
    \item The models can only explain the observations (${\sim} 2$ orders of magnitude spread in line intensities) if both the disk shadowing and dust optical depth effects (i.e. dust attenuation in the envelope, continuum over-subtraction in the disk and larger photodissociation regions) are taken into account (see Fig. \ref{fig:comp_obs}). Considering only one of these effects is not enough to explain the data.
    \item We suggest that most of the objects observed by the PEACHES survey may be hosting a disk and those without strong CH$_3$OH emission are prime targets to search for disks in the embedded phase of star formation.
\end{itemize}

The presence of a disk along with dust optical depth effects can explain the lack of emission in some protostars. Hence, less methanol/complex organic molecule emission does not necessarily imply absence of those molecules or reprocessing of them in the gas phase. This work shows that to understand the chemistry of protostellar systems, the physical structure should be taken into account. In particular, protostars are not necessarily envelope-only objects and the presence of a disk already has an effect on the molecular emission even in the earliest phases of star formation.   

\begin{acknowledgements}
We would like to thank the referee for very useful comments. We would like to thank Jes Jørgensen for helpful discussions. Astrochemistry in Leiden is supported by the Netherlands Research School for Astronomy (NOVA), by funding from the European Research Council (ERC) under the European Union’s Horizon 2020 research and innovation programme (grant agreement No. 101019751 MOLDISK), and by the Dutch Research Council (NWO) grants 648.000.022, 618.000.001 and TOP-1 614.001.751. Support by the Danish National Research Foundation through the Center of Excellence “InterCat” (Grant agreement no.: DNRF150) is also acknowledged. G.R. acknowledges support from the Netherlands Organisation for Scientific Research (NWO, program number 016.Veni.192.233) and from an STFC Ernest Rutherford Fellowship (grant number ST/T003855/1).
\end{acknowledgements}

\bibliographystyle{aa}
\bibliography{env_plus_disk}

\begin{thebibliography}{102}
\expandafter\ifx\csname natexlab\endcsname\relax\def\natexlab#1{#1}\fi

\bibitem[{{Aikawa} {et~al.}(2020){Aikawa}, {Furuya}, {Yamamoto}, \&
  {Sakai}}]{Aikawa2020}
{Aikawa}, Y., {Furuya}, K., {Yamamoto}, S., \& {Sakai}, N. 2020, \apj, 897, 110

\bibitem[{{Artur de la Villarmois} {et~al.}(2019){Artur de la Villarmois},
  {J{\o}rgensen}, {Kristensen}, {Bergin}, {Harsono}, {Sakai}, {van Dishoeck},
  \& {Yamamoto}}]{ArturdelaVillarmois2019}
{Artur de la Villarmois}, E., {J{\o}rgensen}, J.~K., {Kristensen}, L.~E.,
  {et~al.} 2019, \aap, 626, A71

\bibitem[{{Bachiller} \& {Tafalla}(1999)}]{Bachiller1999}
{Bachiller}, R. \& {Tafalla}, M. 1999, in NATO Advanced Study Institute (ASI)
  Series C, Vol. 540, The Origin of Stars and Planetary Systems, ed. C.~J.
  {Lada} \& N.~D. {Kylafis}, 227

\bibitem[{{Belloche} {et~al.}(2020){Belloche}, {Maury}, {Maret}, {Anderl},
  {Bacmann}, {Andr{\'e}}, {Bontemps}, {Cabrit}, {Codella}, {Gaudel}, {Gueth},
  {Lef{\`e}vre}, {Lefloch}, {Podio}, \& {Testi}}]{Belloche2020}
{Belloche}, A., {Maury}, A.~J., {Maret}, S., {et~al.} 2020, \aap, 635, A198

\bibitem[{{Belloche} {et~al.}(2013){Belloche}, {M{\"u}ller}, {Menten},
  {Schilke}, \& {Comito}}]{Belloche2013}
{Belloche}, A., {M{\"u}ller}, H.~S.~P., {Menten}, K.~M., {Schilke}, P., \&
  {Comito}, C. 2013, \aap, 559, A47

\bibitem[{{Bisschop} {et~al.}(2007){Bisschop}, {J{\o}rgensen}, {van Dishoeck},
  \& {de Wachter}}]{Bisschop2007}
{Bisschop}, S.~E., {J{\o}rgensen}, J.~K., {van Dishoeck}, E.~F., \& {de
  Wachter}, E.~B.~M. 2007, \aap, 465, 913

\bibitem[{{Blake} {et~al.}(1987){Blake}, {Sutton}, {Masson}, \&
  {Phillips}}]{Blake1987}
{Blake}, G.~A., {Sutton}, E.~C., {Masson}, C.~R., \& {Phillips}, T.~G. 1987,
  \apj, 315, 621

\bibitem[{{Boehler} {et~al.}(2017){Boehler}, {Weaver}, {Isella}, {Ricci},
  {Grady}, {Carpenter}, \& {Perez}}]{Boehler2017}
{Boehler}, Y., {Weaver}, E., {Isella}, A., {et~al.} 2017, \apj, 840, 60

\bibitem[{{B{\o}gelund} {et~al.}(2018){B{\o}gelund}, {McGuire}, {Ligterink},
  {Taquet}, {Brogan}, {Hunter}, {Pearson}, {Hogerheijde}, \& {van
  Dishoeck}}]{Bogelund2018}
{B{\o}gelund}, E.~G., {McGuire}, B.~A., {Ligterink}, N. F.~W., {et~al.} 2018,
  \aap, 615, A88

\bibitem[{{Boogert} {et~al.}(2015){Boogert}, {Gerakines}, \&
  {Whittet}}]{Boogert2015}
{Boogert}, A.~C.~A., {Gerakines}, P.~A., \& {Whittet}, D. C.~B. 2015, \araa,
  53, 541

\bibitem[{{Boogert} {et~al.}(2011){Boogert}, {Huard}, {Cook}, {Chiar}, {Knez},
  {Decin}, {Blake}, {Tielens}, \& {van Dishoeck}}]{Boogert2011}
{Boogert}, A.~C.~A., {Huard}, T.~L., {Cook}, A.~M., {et~al.} 2011, \apj, 729,
  92

\bibitem[{{Boogert} {et~al.}(2008){Boogert}, {Pontoppidan}, {Knez}, {Lahuis},
  {Kessler-Silacci}, {van Dishoeck}, {Blake}, {Augereau}, {Bisschop},
  {Bottinelli}, {Brooke}, {Brown}, {Crapsi}, {Evans}, {Fraser}, {Geers},
  {Huard}, {J{\o}rgensen}, {{\"O}berg}, {Allen}, {Harvey}, {Koerner}, {Mundy},
  {Padgett}, {Sargent}, \& {Stapelfeldt}}]{Boogert2008}
{Boogert}, A.~C.~A., {Pontoppidan}, K.~M., {Knez}, C., {et~al.} 2008, \apj,
  678, 985

\bibitem[{{Booth} {et~al.}(2021){Booth}, {Walsh}, {Terwisscha van Scheltinga},
  {van Dishoeck}, {Ilee}, {Hogerheijde}, {Kama}, \& {Nomura}}]{Booth2021}
{Booth}, A.~S., {Walsh}, C., {Terwisscha van Scheltinga}, J., {et~al.} 2021,
  Nature Astronomy, 5, 684

\bibitem[{{Bottinelli} {et~al.}(2010){Bottinelli}, {Boogert}, {Bouwman},
  {Beckwith}, {van Dishoeck}, {{\"O}berg}, {Pontoppidan}, {Linnartz}, {Blake},
  {Evans}, \& {Lahuis}}]{Bottinelli2010}
{Bottinelli}, S., {Boogert}, A.~C.~A., {Bouwman}, J., {et~al.} 2010, \apj, 718,
  1100

\bibitem[{{Caselli} \& {Ceccarelli}(2012)}]{Caselli2012}
{Caselli}, P. \& {Ceccarelli}, C. 2012, \aapr, 20, 56

\bibitem[{{Chiang} \& {Goldreich}(1997)}]{Chiang1997}
{Chiang}, E.~I. \& {Goldreich}, P. 1997, \apj, 490, 368

\bibitem[{{Crapsi} {et~al.}(2008){Crapsi}, {van Dishoeck}, {Hogerheijde},
  {Pontoppidan}, \& {Dullemond}}]{Crapsi2008}
{Crapsi}, A., {van Dishoeck}, E.~F., {Hogerheijde}, M.~R., {Pontoppidan},
  K.~M., \& {Dullemond}, C.~P. 2008, \aap, 486, 245

\bibitem[{{Dartois} {et~al.}(2002){Dartois}, {d'Hendecourt}, {Thi},
  {Pontoppidan}, \& {van Dishoeck}}]{Dartois2002}
{Dartois}, E., {d'Hendecourt}, L., {Thi}, W., {Pontoppidan}, K.~M., \& {van
  Dishoeck}, E.~F. 2002, \aap, 394, 1057

\bibitem[{{Dartois} {et~al.}(1999){Dartois}, {Schutte}, {Geballe}, {Demyk},
  {Ehrenfreund}, \& {D'Hendecourt}}]{Dartois1999}
{Dartois}, E., {Schutte}, W., {Geballe}, T.~R., {et~al.} 1999, \aap, 342, L32

\bibitem[{{De Simone} {et~al.}(2020){De Simone}, {Ceccarelli}, {Codella},
  {Svoboda}, {Chandler}, {Bouvier}, {Yamamoto}, {Sakai}, {Caselli}, {Favre},
  {Loinard}, {Lefloch}, {Liu}, {L{\'o}pez-Sepulcre}, {Pineda}, {Taquet}, \&
  {Testi}}]{DeSimone2020}
{De Simone}, M., {Ceccarelli}, C., {Codella}, C., {et~al.} 2020, \apjl, 896, L3

\bibitem[{{Draine}(2011)}]{Draine2011}
{Draine}, B.~T. 2011, {Physics of the Interstellar and Intergalactic Medium}

\bibitem[{{Drozdovskaya} {et~al.}(2014){Drozdovskaya}, {Walsh}, {Visser},
  {Harsono}, \& {van Dishoeck}}]{Drozdovskaya2014}
{Drozdovskaya}, M.~N., {Walsh}, C., {Visser}, R., {Harsono}, D., \& {van
  Dishoeck}, E.~F. 2014, \mnras, 445, 913

\bibitem[{{Drozdovskaya} {et~al.}(2015){Drozdovskaya}, {Walsh}, {Visser},
  {Harsono}, \& {van Dishoeck}}]{Drozdovskaya2015}
{Drozdovskaya}, M.~N., {Walsh}, C., {Visser}, R., {Harsono}, D., \& {van
  Dishoeck}, E.~F. 2015, \mnras, 451, 3836

\bibitem[{{Dullemond} {et~al.}(2001){Dullemond}, {Dominik}, \&
  {Natta}}]{Dullemond2001}
{Dullemond}, C.~P., {Dominik}, C., \& {Natta}, A. 2001, \apj, 560, 957

\bibitem[{{Forbrich} {et~al.}(2006){Forbrich}, {Preibisch}, \&
  {Menten}}]{Forbrich2006}
{Forbrich}, J., {Preibisch}, T., \& {Menten}, K.~M. 2006, \aap, 446, 155

\bibitem[{{Fuchs} {et~al.}(2009){Fuchs}, {Cuppen}, {Ioppolo}, {Romanzin},
  {Bisschop}, {Andersson}, {van Dishoeck}, \& {Linnartz}}]{Fuchs2009}
{Fuchs}, G.~W., {Cuppen}, H.~M., {Ioppolo}, S., {et~al.} 2009, \aap, 505, 629

\bibitem[{{Garrod} \& {Pauly}(2011)}]{Garrod2011}
{Garrod}, R.~T. \& {Pauly}, T. 2011, \apj, 735, 15

\bibitem[{{Gavino} {et~al.}(2021){Gavino}, {Dutrey}, {Wakelam}, {Guilloteau},
  {Kobus}, {Wolf}, {Iqbal}, {Di Folco}, {Chapillon}, \&
  {Pi{\'e}tu}}]{Gavino2021}
{Gavino}, S., {Dutrey}, A., {Wakelam}, V., {et~al.} 2021, arXiv e-prints,
  arXiv:2106.05888

\bibitem[{{Geballe} {et~al.}(1988){Geballe}, {Kim}, {Knacke}, \&
  {Noll}}]{Geballe1988}
{Geballe}, T.~R., {Kim}, Y.~H., {Knacke}, R.~F., \& {Noll}, K.~S. 1988, \apjl,
  326, L65

\bibitem[{{Geppert} {et~al.}(2006){Geppert}, {Hamberg}, {Thomas},
  {{\"O}sterdahl}, {Hellberg}, {Zhaunerchyk}, {Ehlerding}, {Millar}, {Roberts},
  {Semaniak}, {Ugglas}, {K{\"a}llberg}, {Simonsson}, {Kaminska}, \&
  {Larsson}}]{Geppert2006}
{Geppert}, W.~D., {Hamberg}, M., {Thomas}, R.~D., {et~al.} 2006, Faraday
  Discussions, 133, 177

\bibitem[{{Gibb} {et~al.}(2000){Gibb}, {Nummelin}, {Irvine}, {Whittet}, \&
  {Bergman}}]{Gibb2000}
{Gibb}, E., {Nummelin}, A., {Irvine}, W.~M., {Whittet}, D.~C.~B., \& {Bergman},
  P. 2000, \apj, 545, 309

\bibitem[{{Grosso} {et~al.}(2020){Grosso}, {Hamaguchi}, {Principe}, \&
  {Kastner}}]{Grosso2020}
{Grosso}, N., {Hamaguchi}, K., {Principe}, D.~A., \& {Kastner}, J.~H. 2020,
  \aap, 638, L4

\bibitem[{{Harsono} {et~al.}(2015){Harsono}, {Bruderer}, \& {van
  Dishoeck}}]{Harsono2015}
{Harsono}, D., {Bruderer}, S., \& {van Dishoeck}, E.~F. 2015, \aap, 582, A41

\bibitem[{{Hasegawa} {et~al.}(1992){Hasegawa}, {Herbst}, \&
  {Leung}}]{Hasegawa1992}
{Hasegawa}, T.~I., {Herbst}, E., \& {Leung}, C.~M. 1992, \apjs, 82, 167

\bibitem[{{Heays} {et~al.}(2017){Heays}, {Bosman}, \& {van
  Dishoeck}}]{Heays2017}
{Heays}, A.~N., {Bosman}, A.~D., \& {van Dishoeck}, E.~F. 2017, \aap, 602, A105

\bibitem[{{Herbst} \& {van Dishoeck}(2009)}]{Herbst2009}
{Herbst}, E. \& {van Dishoeck}, E.~F. 2009, \araa, 47, 427

\bibitem[{{Hidaka} {et~al.}(2004){Hidaka}, {Watanabe}, {Shiraki}, {Nagaoka}, \&
  {Kouchi}}]{Hidaka2004}
{Hidaka}, H., {Watanabe}, N., {Shiraki}, T., {Nagaoka}, A., \& {Kouchi}, A.
  2004, \apj, 614, 1124

\bibitem[{{Hogerheijde} \& {van der Tak}(2000)}]{2000A&A...362..697H}
{Hogerheijde}, M.~R. \& {van der Tak}, F.~F.~S. 2000, \aap, 362, 697

\bibitem[{{Ilee} {et~al.}(2016){Ilee}, {Cyganowski}, {Nazari}, {Hunter},
  {Brogan}, {Forgan}, \& {Zhang}}]{Ilee2016}
{Ilee}, J.~D., {Cyganowski}, C.~J., {Nazari}, P., {et~al.} 2016, \mnras, 462,
  4386

\bibitem[{{J{\o}rgensen} {et~al.}(2002){J{\o}rgensen}, {Sch{\"o}ier}, \& {van
  Dishoeck}}]{Jorgensen2002}
{J{\o}rgensen}, J.~K., {Sch{\"o}ier}, F.~L., \& {van Dishoeck}, E.~F. 2002,
  \aap, 389, 908

\bibitem[{{J{\o}rgensen} {et~al.}(2005){J{\o}rgensen}, {Sch{\"o}ier}, \& {van
  Dishoeck}}]{Jorgensen2005}
{J{\o}rgensen}, J.~K., {Sch{\"o}ier}, F.~L., \& {van Dishoeck}, E.~F. 2005,
  \aap, 437, 501

\bibitem[{{J{\o}rgensen} {et~al.}(2016){J{\o}rgensen}, {van der Wiel},
  {Coutens}, {Lykke}, {M{\"u}ller}, {van Dishoeck}, {Calcutt}, {Bjerkeli},
  {Bourke}, {Drozdovskaya}, {Favre}, {Fayolle}, {Garrod}, {Jacobsen},
  {{\"O}berg}, {Persson}, \& {Wampfler}}]{Jorgensen2016}
{J{\o}rgensen}, J.~K., {van der Wiel}, M.~H.~D., {Coutens}, A., {et~al.} 2016,
  \aap, 595, A117

\bibitem[{{J{\o}rgensen} {et~al.}(2009){J{\o}rgensen}, {van Dishoeck},
  {Visser}, {Bourke}, {Wilner}, {Lommen}, {Hogerheijde}, \&
  {Myers}}]{Jorgensen2009}
{J{\o}rgensen}, J.~K., {van Dishoeck}, E.~F., {Visser}, R., {et~al.} 2009,
  \aap, 507, 861

\bibitem[{{Kristensen} {et~al.}(2012){Kristensen}, {van Dishoeck}, {Bergin},
  {Visser}, {Y{\i}ld{\i}z}, {San Jose-Garcia}, {J{\o}rgensen}, {Herczeg},
  {Johnstone}, {Wampfler}, {Benz}, {Bruderer}, {Cabrit}, {Caselli}, {Doty},
  {Harsono}, {Herpin}, {Hogerheijde}, {Karska}, {van Kempen}, {Liseau},
  {Nisini}, {Tafalla}, {van der Tak}, \& {Wyrowski}}]{Kristensen2012}
{Kristensen}, L.~E., {van Dishoeck}, E.~F., {Bergin}, E.~A., {et~al.} 2012,
  \aap, 542, A8

\bibitem[{{Laas} {et~al.}(2011){Laas}, {Garrod}, {Herbst}, \& {Widicus
  Weaver}}]{Laas2011}
{Laas}, J.~C., {Garrod}, R.~T., {Herbst}, E., \& {Widicus Weaver}, S.~L. 2011,
  \apj, 728, 71

\bibitem[{{Law} {et~al.}(2021){Law}, {Zhang}, {{\"O}berg}, {Galv{\'a}n-Madrid},
  {Keto}, {Liu}, \& {Ho}}]{Law2021}
{Law}, C.~J., {Zhang}, Q., {{\"O}berg}, K.~I., {et~al.} 2021, \apj, 909, 214

\bibitem[{{Lee} {et~al.}(2019){Lee}, {Codella}, {Li}, \& {Liu}}]{Lee2019}
{Lee}, C.-F., {Codella}, C., {Li}, Z.-Y., \& {Liu}, S.-Y. 2019, \apj, 876, 63

\bibitem[{{Lee} {et~al.}(2017){Lee}, {Li}, {Ho}, {Hirano}, {Zhang}, \&
  {Shang}}]{Lee2017}
{Lee}, C.-F., {Li}, Z.-Y., {Ho}, P. T.~P., {et~al.} 2017, \apj, 843, 27

\bibitem[{{Ligterink} {et~al.}(2021){Ligterink}, {Ahmadi}, {Coutens},
  {Tychoniec}, {Calcutt}, {van Dishoeck}, {Linnartz}, {J{\o}rgensen}, {Garrod},
  \& {Bouwman}}]{Ligterink2021}
{Ligterink}, N.~F.~W., {Ahmadi}, A., {Coutens}, A., {et~al.} 2021, \aap, 647,
  A87

\bibitem[{{L{\'o}pez-Sepulcre} {et~al.}(2017){L{\'o}pez-Sepulcre}, {Sakai},
  {Neri}, {Imai}, {Oya}, {Ceccarelli}, {Higuchi}, {Aikawa}, {Bottinelli},
  {Caux}, {Hirota}, {Kahane}, {Lefloch}, {Vastel}, {Watanabe}, \&
  {Yamamoto}}]{Lopez-spulcre2017}
{L{\'o}pez-Sepulcre}, A., {Sakai}, N., {Neri}, R., {et~al.} 2017, \aap, 606,
  A121

\bibitem[{{Manigand} {et~al.}(2020){Manigand}, {J{\o}rgensen}, {Calcutt},
  {M{\"u}ller}, {Ligterink}, {Coutens}, {Drozdovskaya}, {van Dishoeck}, \&
  {Wampfler}}]{Manigand2020}
{Manigand}, S., {J{\o}rgensen}, J.~K., {Calcutt}, H., {et~al.} 2020, \aap, 635,
  A48

\bibitem[{{Marcelino} {et~al.}(2018){Marcelino}, {Gerin}, {Cernicharo},
  {Fuente}, {Wootten}, {Chapillon}, {Pety}, {Lis}, {Roueff}, {Commer{\c{c}}on},
  \& {Ciardi}}]{Marcelino2018}
{Marcelino}, N., {Gerin}, M., {Cernicharo}, J., {et~al.} 2018, \aap, 620, A80

\bibitem[{{Maret} {et~al.}(2004){Maret}, {Ceccarelli}, {Caux}, {Tielens},
  {J{\o}rgensen}, {van Dishoeck}, {Bacmann}, {Castets}, {Lefloch}, {Loinard},
  {Parise}, \& {Sch{\"o}ier}}]{Maret2004}
{Maret}, S., {Ceccarelli}, C., {Caux}, E., {et~al.} 2004, \aap, 416, 577

\bibitem[{{Maret} {et~al.}(2005){Maret}, {Ceccarelli}, {Tielens}, {Caux},
  {Lefloch}, {Faure}, {Castets}, \& {Flower}}]{Maret2005}
{Maret}, S., {Ceccarelli}, C., {Tielens}, A.~G.~G.~M., {et~al.} 2005, \aap,
  442, 527

\bibitem[{{Mart{\'\i}n-Dom{\'e}nech} {et~al.}(2019){Mart{\'\i}n-Dom{\'e}nech},
  {Bergner}, {{\"O}berg}, \& {J{\o}rgensen}}]{Martin-Domenech2019}
{Mart{\'\i}n-Dom{\'e}nech}, R., {Bergner}, J.~B., {{\"O}berg}, K.~I., \&
  {J{\o}rgensen}, J.~K. 2019, \apj, 880, 130

\bibitem[{{Maury} {et~al.}(2019){Maury}, {Andr{\'e}}, {Testi}, {Maret},
  {Belloche}, {Hennebelle}, {Cabrit}, {Codella}, {Gueth}, {Podio}, {Anderl},
  {Bacmann}, {Bontemps}, {Gaudel}, {Ladjelate}, {Lef{\`e}vre}, {Tabone}, \&
  {Lefloch}}]{Maury2019}
{Maury}, A.~J., {Andr{\'e}}, P., {Testi}, L., {et~al.} 2019, \aap, 621, A76

\bibitem[{{McGuire} {et~al.}(2017){McGuire}, {Shingledecker}, {Willis},
  {Burkhardt}, {El-Abd}, {Motiyenko}, {Brogan}, {Hunter}, {Margul{\`e}s},
  {Guillemin}, {Garrod}, {Herbst}, \& {Remijan}}]{McGuire2017}
{McGuire}, B.~A., {Shingledecker}, C.~N., {Willis}, E.~R., {et~al.} 2017,
  \apjl, 851, L46

\bibitem[{{Mottram} {et~al.}(2017){Mottram}, {van Dishoeck}, {Kristensen},
  {Karska}, {San Jos{\'e}-Garc{\'\i}a}, {Khanna}, {Herczeg}, {Andr{\'e}},
  {Bontemps}, {Cabrit}, {Carney}, {Drozdovskaya}, {Dunham}, {Evans}, {Fedele},
  {Green}, {Harsono}, {Johnstone}, {J{\o}rgensen}, {K{\"o}nyves}, {Nisini},
  {Persson}, {Tafalla}, {Visser}, \& {Y{\i}ld{\i}z}}]{Mottram2017}
{Mottram}, J.~C., {van Dishoeck}, E.~F., {Kristensen}, L.~E., {et~al.} 2017,
  \aap, 600, A99

\bibitem[{{M{\"u}ller} {et~al.}(2005){M{\"u}ller}, {Schl{\"o}der}, {Stutzki},
  \& {Winnewisser}}]{Muller2005}
{M{\"u}ller}, H. S.~P., {Schl{\"o}der}, F., {Stutzki}, J., \& {Winnewisser}, G.
  2005, Journal of Molecular Structure, 742, 215

\bibitem[{{M{\"u}ller} {et~al.}(2001){M{\"u}ller}, {Thorwirth}, {Roth}, \&
  {Winnewisser}}]{Muller2001}
{M{\"u}ller}, H.~S.~P., {Thorwirth}, S., {Roth}, D.~A., \& {Winnewisser}, G.
  2001, \aap, 370, L49

\bibitem[{{Murillo} {et~al.}(2015){Murillo}, {Bruderer}, {van Dishoeck},
  {Walsh}, {Harsono}, {Lai}, \& {Fuchs}}]{Murillo2015}
{Murillo}, N.~M., {Bruderer}, S., {van Dishoeck}, E.~F., {et~al.} 2015, \aap,
  579, A114

\bibitem[{{Murillo} {et~al.}(2013){Murillo}, {Lai}, {Bruderer}, {Harsono}, \&
  {van Dishoeck}}]{Murillo2013}
{Murillo}, N.~M., {Lai}, S.-P., {Bruderer}, S., {Harsono}, D., \& {van
  Dishoeck}, E.~F. 2013, \aap, 560, A103

\bibitem[{{Murillo} {et~al.}(2018){Murillo}, {van Dishoeck}, {van der Wiel},
  {J{\o}rgensen}, {Drozdovskaya}, {Calcutt}, \& {Harsono}}]{Murillo2018}
{Murillo}, N.~M., {van Dishoeck}, E.~F., {van der Wiel}, M.~H.~D., {et~al.}
  2018, \aap, 617, A120

\bibitem[{{Nazari} {et~al.}(2021){Nazari}, {van Gelder}, {van Dishoeck},
  {Tabone}, {van't Hoff}, {Ligterink}, {Beuther}, {Boogert}, {Caratti o
  Garatti}, {Klaassen}, {Linnartz}, {Taquet}, \& {Tychoniec}}]{Nazari2021}
{Nazari}, P., {van Gelder}, M.~L., {van Dishoeck}, E.~F., {et~al.} 2021, \aap,
  650, A150

\bibitem[{{Notsu} {et~al.}(2021){Notsu}, {van Dishoeck}, {Walsh}, {Bosman}, \&
  {Nomura}}]{Notsu2021}
{Notsu}, S., {van Dishoeck}, E.~F., {Walsh}, C., {Bosman}, A.~D., \& {Nomura},
  H. 2021, \aap, 650, A180

\bibitem[{{Nourry} \& {Krim}(2015)}]{Nourry2015}
{Nourry}, S. \& {Krim}, L. 2015, \mnras, 452, 3319

\bibitem[{{{\"O}berg} {et~al.}(2011){{\"O}berg}, {Boogert}, {Pontoppidan}, {van
  den Broek}, {van Dishoeck}, {Bottinelli}, {Blake}, \& {Evans}}]{Oberg2011}
{{\"O}berg}, K.~I., {Boogert}, A.~C.~A., {Pontoppidan}, K.~M., {et~al.} 2011,
  \apj, 740, 109

\bibitem[{{Penteado} {et~al.}(2017){Penteado}, {Walsh}, \&
  {Cuppen}}]{Penteado2017}
{Penteado}, E.~M., {Walsh}, C., \& {Cuppen}, H.~M. 2017, \apj, 844, 71

\bibitem[{{Persson} {et~al.}(2016){Persson}, {Harsono}, {Tobin}, {van
  Dishoeck}, {J{\o}rgensen}, {Murillo}, \& {Lai}}]{Persson2016}
{Persson}, M.~V., {Harsono}, D., {Tobin}, J.~J., {et~al.} 2016, \aap, 590, A33

\bibitem[{{Persson} {et~al.}(2012){Persson}, {J{\o}rgensen}, \& {van
  Dishoeck}}]{Persson2012}
{Persson}, M.~V., {J{\o}rgensen}, J.~K., \& {van Dishoeck}, E.~F. 2012, \aap,
  541, A39

\bibitem[{{Plunkett} {et~al.}(2013){Plunkett}, {Arce}, {Corder}, {Mardones},
  {Sargent}, \& {Schnee}}]{Plunkett2013}
{Plunkett}, A.~L., {Arce}, H.~G., {Corder}, S.~A., {et~al.} 2013, \apj, 774, 22

\bibitem[{{Podio} {et~al.}(2020){Podio}, {Garufi}, {Codella}, {Fedele},
  {Bianchi}, {Bacciotti}, {Ceccarelli}, {Favre}, {Mercimek}, {Rygl}, \&
  {Testi}}]{Podio2020}
{Podio}, L., {Garufi}, A., {Codella}, C., {et~al.} 2020, \aap, 642, L7

\bibitem[{{Pringle}(1981)}]{Pringle1981}
{Pringle}, J.~E. 1981, \araa, 19, 137

\bibitem[{{Rabli} \& {Flower}(2010{\natexlab{a}})}]{Rabli2010}
{Rabli}, D. \& {Flower}, D.~R. 2010{\natexlab{a}}, \mnras, 406, 95

\bibitem[{{Rabli} \& {Flower}(2010{\natexlab{b}})}]{2010MNRAS.406...95R}
{Rabli}, D. \& {Flower}, D.~R. 2010{\natexlab{b}}, \mnras, 406, 95

\bibitem[{{Rosotti} {et~al.}(2021){Rosotti}, {Ilee}, {Facchini}, {Tazzari},
  {Booth}, {Clarke}, \& {Kama}}]{Rosotti2021}
{Rosotti}, G.~P., {Ilee}, J.~D., {Facchini}, S., {et~al.} 2021, \mnras, 501,
  3427

\bibitem[{{Sanhueza} {et~al.}(2013){Sanhueza}, {Jackson}, {Foster},
  {Jimenez-Serra}, {Dirienzo}, \& {Pillai}}]{Sanhueza2013}
{Sanhueza}, P., {Jackson}, J.~M., {Foster}, J.~B., {et~al.} 2013, \apj, 773,
  123

\bibitem[{{Sch{\"o}ier} {et~al.}(2005){Sch{\"o}ier}, {van der Tak}, {van
  Dishoeck}, \& {Black}}]{Schoier2005}
{Sch{\"o}ier}, F.~L., {van der Tak}, F.~F.~S., {van Dishoeck}, E.~F., \&
  {Black}, J.~H. 2005, \aap, 432, 369

\bibitem[{{Scibelli} {et~al.}(2021){Scibelli}, {Shirley}, {Vasyunin}, \&
  {Launhardt}}]{Scibelli2021}
{Scibelli}, S., {Shirley}, Y., {Vasyunin}, A., \& {Launhardt}, R. 2021, \mnras,
  504, 5754

\bibitem[{{Segura-Cox} {et~al.}(2018){Segura-Cox}, {Looney}, {Tobin}, {Li},
  {Harris}, {Sadavoy}, {Dunham}, {Chandler}, {Kratter}, {P{\'e}rez}, \&
  {Melis}}]{Segura-Cox2018}
{Segura-Cox}, D.~M., {Looney}, L.~W., {Tobin}, J.~J., {et~al.} 2018, \apj, 866,
  161

\bibitem[{{Shakura} \& {Sunyaev}(1973)}]{Shakura1973}
{Shakura}, N.~I. \& {Sunyaev}, R.~A. 1973, \aap, 500, 33

\bibitem[{{Tabone} {et~al.}(2021){Tabone}, {van Hemert}, {van Dishoeck}, \&
  {Black}}]{Tabone2021}
{Tabone}, B., {van Hemert}, M.~C., {van Dishoeck}, E.~F., \& {Black}, J.~H.
  2021, \aap, 650, A192

\bibitem[{{Taquet} {et~al.}(2019){Taquet}, {Bianchi}, {Codella}, {Persson},
  {Ceccarelli}, {Cabrit}, {J{\o}rgensen}, {Kahane}, {L{\'o}pez-Sepulcre}, \&
  {Neri}}]{Taquet2019}
{Taquet}, V., {Bianchi}, E., {Codella}, C., {et~al.} 2019, \aap, 632, A19

\bibitem[{{Tobin} {et~al.}(2020){Tobin}, {Sheehan}, {Megeath},
  {D{\'\i}az-Rodr{\'\i}guez}, {Offner}, {Murillo}, {van 't Hoff}, {van
  Dishoeck}, {Osorio}, {Anglada}, {Furlan}, {Stutz}, {Reynolds}, {Karnath},
  {Fischer}, {Persson}, {Looney}, {Li}, {Stephens}, {Chandler}, {Cox},
  {Dunham}, {Tychoniec}, {Kama}, {Kratter}, {Kounkel}, {Mazur}, {Maud},
  {Patel}, {Perez}, {Sadavoy}, {Segura-Cox}, {Sharma}, {Stephenson}, {Watson},
  \& {Wyrowski}}]{Tobin2020}
{Tobin}, J.~J., {Sheehan}, P.~D., {Megeath}, S.~T., {et~al.} 2020, \apj, 890,
  130

\bibitem[{{Tychoniec} {et~al.}(2021){Tychoniec}, {van Dishoeck}, {van 't Hoff},
  {van Gelder}, {Tabone}, {Chen}, {Harsono}, {Hull}, {Hogerheijde}, {Murillo},
  \& {Tobin}}]{Tychoniec2021}
{Tychoniec}, {\L}., {van Dishoeck}, E.~F., {van 't Hoff}, M. L.~R., {et~al.}
  2021, arXiv e-prints, arXiv:2107.03696

\bibitem[{{Ulrich}(1976)}]{Ulrich1976}
{Ulrich}, R.~K. 1976, \apj, 210, 377

\bibitem[{{van Dishoeck} {et~al.}(1995){van Dishoeck}, {Blake}, {Jansen}, \&
  {Groesbeck}}]{Ewine1995}
{van Dishoeck}, E.~F., {Blake}, G.~A., {Jansen}, D.~J., \& {Groesbeck}, T.~D.
  1995, \apj, 447, 760

\bibitem[{{van Dishoeck} {et~al.}(2021){van Dishoeck}, {Kristensen}, {Mottram},
  {Benz}, {Bergin}, {Caselli}, {Herpin}, {Hogerheijde}, {Johnstone}, {Liseau},
  {Nisini}, {Tafalla}, {van der Tak}, {Wyrowski}, {Baudry}, {Benedettini},
  {Bjerkeli}, {Blake}, {Braine}, {Bruderer}, {Cabrit}, {Cernicharo}, {Choi},
  {Coutens}, {de Graauw}, {Dominik}, {Fedele}, {Fich}, {Fuente}, {Furuya},
  {Goicoechea}, {Harsono}, {Helmich}, {Herczeg}, {Jacq}, {Karska}, {Kaufman},
  {Keto}, {Lamberts}, {Larsson}, {Leurini}, {Lis}, {Melnick}, {Neufeld},
  {Pagani}, {Persson}, {Shipman}, {Taquet}, {van Kempen}, {Walsh}, {Wampfler},
  {Y{\i}ld{\i}z}, \& {WISH Team}}]{vanDishoeck2021}
{van Dishoeck}, E.~F., {Kristensen}, L.~E., {Mottram}, J.~C., {et~al.} 2021,
  \aap, 648, A24

\bibitem[{{van Gelder} {et~al.}(2022){van Gelder}, {Nazari}, {Tabone},
  {Ahmadi}, {van Dishoeck}, {Beltr{\'a}n}, {Fuller}, {Sakai},
  {S{\'a}nchez-Monge}, {Schilke}, {Yang}, \& {Zhang}}]{vanGelder2022}
{van Gelder}, M.~L., {Nazari}, P., {Tabone}, B., {et~al.} 2022, arXiv e-prints,
  arXiv:2202.04723

\bibitem[{{van Gelder} {et~al.}(2020){van Gelder}, {Tabone}, {Tychoniec}, {van
  Dishoeck}, {Beuther}, {Boogert}, {Caratti o Garatti}, {Klaassen}, {Linnartz},
  {M{\"u}ller}, \& {Taquet}}]{vangelder2020}
{van Gelder}, M.~L., {Tabone}, B., {Tychoniec}, {\L}., {et~al.} 2020, \aap,
  639, A87

\bibitem[{{van Kempen} {et~al.}(2009){van Kempen}, {van Dishoeck},
  {G{\"u}sten}, {Kristensen}, {Schilke}, {Hogerheijde}, {Boland}, {Nefs},
  {Menten}, {Baryshev}, \& {Wyrowski}}]{vanKempen2009}
{van Kempen}, T.~A., {van Dishoeck}, E.~F., {G{\"u}sten}, R., {et~al.} 2009,
  \aap, 501, 633

\bibitem[{{van 't Hoff} {et~al.}(2020){van 't Hoff}, {Harsono}, {Tobin},
  {Bosman}, {van Dishoeck}, {J{\o}rgensen}, {Miotello}, {Murillo}, \&
  {Walsh}}]{vantHoff2020c}
{van 't Hoff}, M. L.~R., {Harsono}, D., {Tobin}, J.~J., {et~al.} 2020, \apj,
  901, 166

\bibitem[{{van 't Hoff} {et~al.}(2021){van 't Hoff}, {Harsono}, {van Gelder},
  {Hsieh}, {Tobin}, {Jensen}, {Hirano}, {J{\o}rgensen}, {Bergin}, \& {van
  Dishoeck}}]{vantHoff2021}
{van 't Hoff}, M. L.~R., {Harsono}, D., {van Gelder}, M.~L., {et~al.} 2021,
  arXiv e-prints, arXiv:2110.08286

\bibitem[{{Visser} {et~al.}(2011){Visser}, {Doty}, \& {van
  Dishoeck}}]{Visser2011}
{Visser}, R., {Doty}, S.~D., \& {van Dishoeck}, E.~F. 2011, \aap, 534, A132

\bibitem[{{Visser} {et~al.}(2013){Visser}, {J{\o}rgensen}, {Kristensen}, {van
  Dishoeck}, \& {Bergin}}]{Visser2013}
{Visser}, R., {J{\o}rgensen}, J.~K., {Kristensen}, L.~E., {van Dishoeck},
  E.~F., \& {Bergin}, E.~A. 2013, \apj, 769, 19

\bibitem[{{Walsh} {et~al.}(2014){Walsh}, {Millar}, {Nomura}, {Herbst}, {Widicus
  Weaver}, {Aikawa}, {Laas}, \& {Vasyunin}}]{Walsh2014}
{Walsh}, C., {Millar}, T.~J., {Nomura}, H., {et~al.} 2014, \aap, 563, A33

\bibitem[{{Weaver} {et~al.}(2018){Weaver}, {Isella}, \& {Boehler}}]{Weaver2018}
{Weaver}, E., {Isella}, A., \& {Boehler}, Y. 2018, \apj, 853, 113

\bibitem[{{Whitney} {et~al.}(2003){Whitney}, {Wood}, {Bjorkman}, \&
  {Wolff}}]{Whitney2003}
{Whitney}, B.~A., {Wood}, K., {Bjorkman}, J.~E., \& {Wolff}, M.~J. 2003, \apj,
  591, 1049

\bibitem[{{Xu} {et~al.}(2008){Xu}, {Fisher}, {Lees}, {Shi}, {Hougen},
  {Pearson}, {Drouin}, {Blake}, \& {Braakman}}]{Xu2008}
{Xu}, L.-H., {Fisher}, J., {Lees}, R.~M., {et~al.} 2008, Journal of Molecular
  Spectroscopy, 251, 305

\bibitem[{{Yang} {et~al.}(2020){Yang}, {Evans}, {Smith}, {Lee}, {Tobin},
  {Terebey}, {Calcutt}, {J{\o}rgensen}, {Green}, \& {Bourke}}]{Yang2020}
{Yang}, Y.-L., {Evans}, Neal~J., I., {Smith}, A., {et~al.} 2020, \apj, 891, 61

\bibitem[{{Yang} {et~al.}(2021){Yang}, {Sakai}, {Zhang}, {Murillo}, {Zhang},
  {Higuchi}, {Zeng}, {L{\'o}pez-Sepulcre}, {Yamamoto}, {Lefloch}, {Bouvier},
  {Ceccarelli}, {Hirota}, {Imai}, {Oya}, {Sakai}, \& {Watanabe}}]{Yang2021}
{Yang}, Y.-L., {Sakai}, N., {Zhang}, Y., {et~al.} 2021, \apj, 910, 20

\bibitem[{{Ysard} {et~al.}(2019){Ysard}, {Koehler}, {Jimenez-Serra}, {Jones},
  \& {Verstraete}}]{Ysard2019}
{Ysard}, N., {Koehler}, M., {Jimenez-Serra}, I., {Jones}, A.~P., \&
  {Verstraete}, L. 2019, \aap, 631, A88

\end{thebibliography}

\begin{appendix} 

\section{Optical depth}
\label{app:opacity_law}

\begin{figure}
  \resizebox{0.9\hsize}{!}{\includegraphics{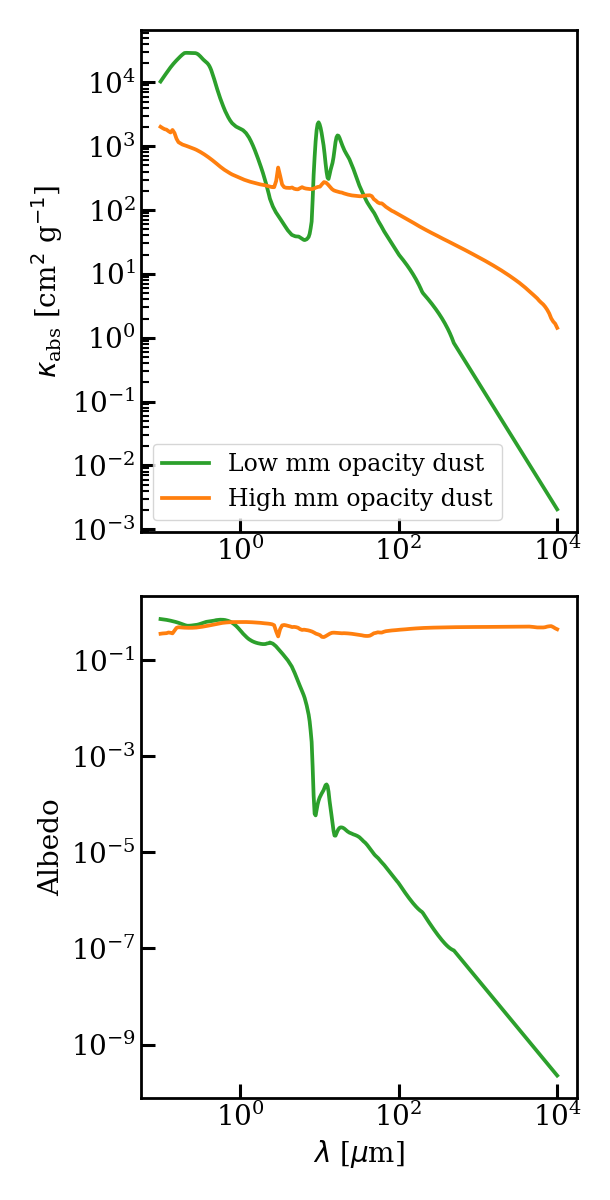}}
  \caption{The properties of the dust grains used in this work. Albedo is defined as the scattering opacity divided by the sum of the absorption and scattering opacities.}
  \label{fig:opacities}
\end{figure}

\begin{figure}
  \resizebox{\hsize}{!}{\includegraphics{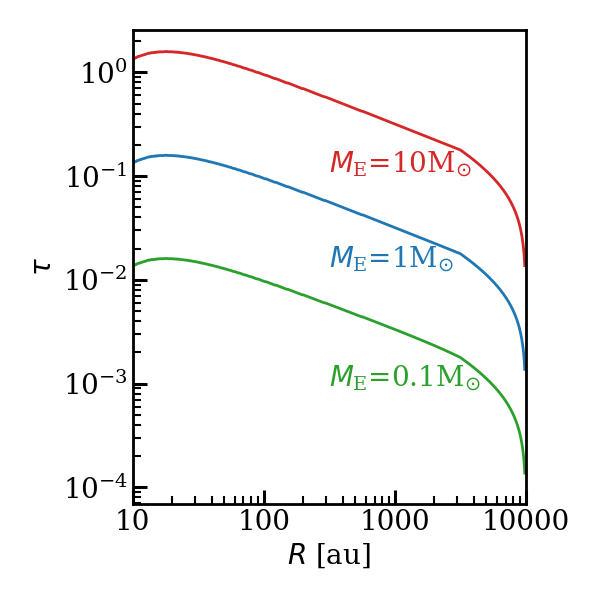}}
  \caption{Radial cut of dust optical depth for the fiducial model with high mm opacity dust in the envelope-only model with different envelope masses. The envelope becomes marginally optically thick in the inner ${\sim}100$\,au for envelope masses above 1\,M$_{\odot}$.}
  \label{fig:tau_fiducial_largedust}
\end{figure}

Figure \ref{fig:opacities} shows the optical properties of the two dust distributions used in this work. The top panel shows the dust absorption opacity and the bottom panel shows the albedo. Figure\ref{fig:tau_fiducial_largedust} shows the dust optical depth for the fiducial envelope-only models with different envelope masses and high mm opacity dust grains.

\begin{figure}
	\resizebox{\hsize}{!}{\includegraphics{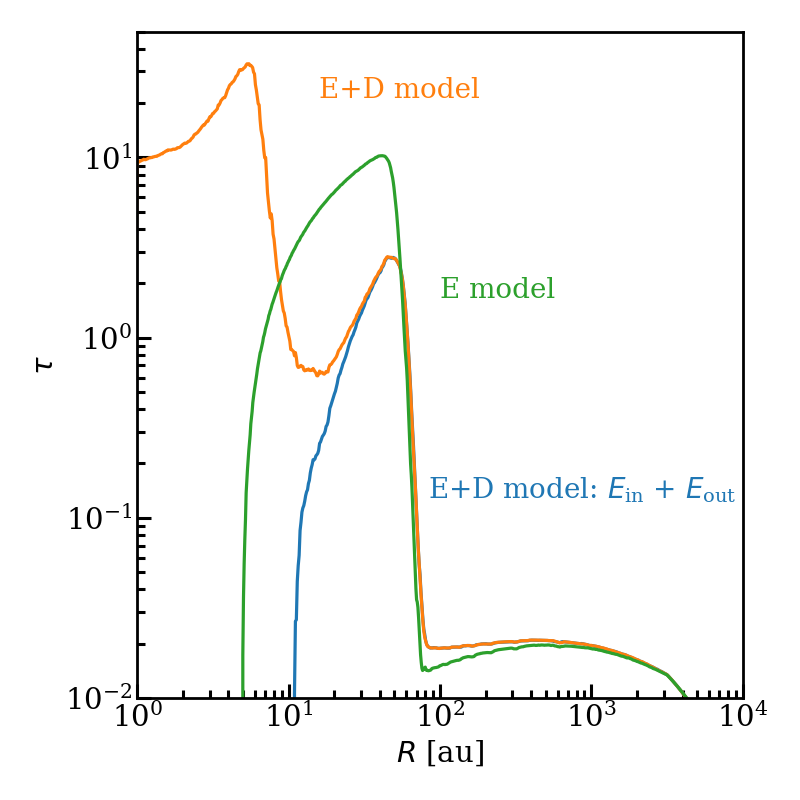}}
	\caption{Radial cut through the line optical depth for the fiducial models (small $\kappa_{\rm dust}$ at mm). Orange and green show $\tau$ for the envelope-plus-disk and envelope-only fiducial models. Blue shows $\tau$ for the envelope component of the envelope-plus-disk model where the model is run by setting methanol abundance to zero in the disk. The line is more optically thick in the disk than the envelope.}
  \label{fig:tau_line}
\end{figure}

\begin{figure}
	\resizebox{\hsize}{!}{\includegraphics{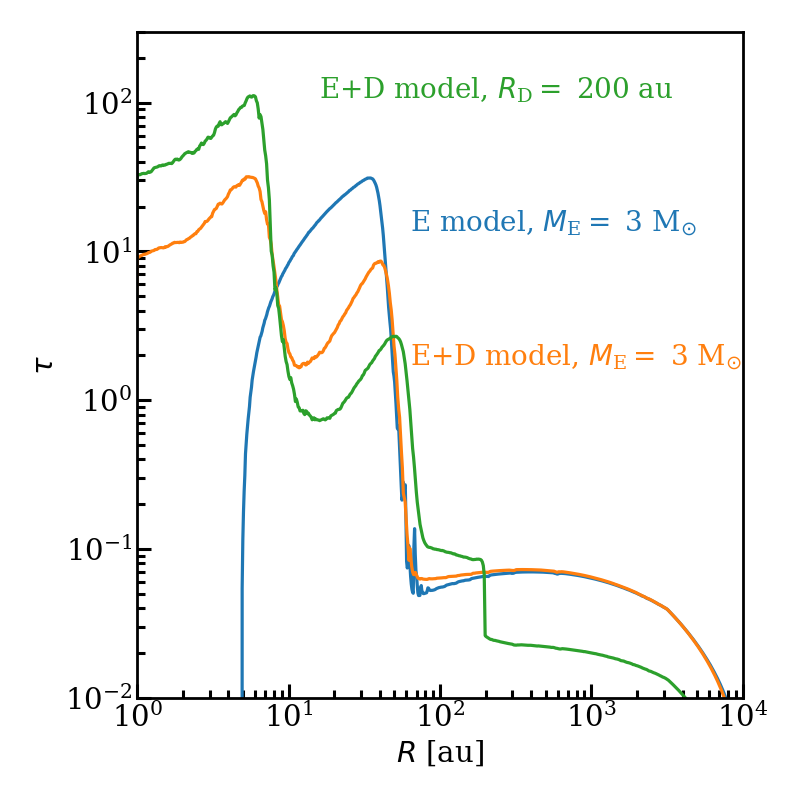}}
	\caption{Radial cut through the line optical depth for fiducial models with envelope mass of 3\,M$_{\odot}$ and fiducial models with disk radius of 200\,au.}
  \label{fig:tau_line_M3}
\end{figure}

\begin{figure}
	\resizebox{\hsize}{!}{\includegraphics{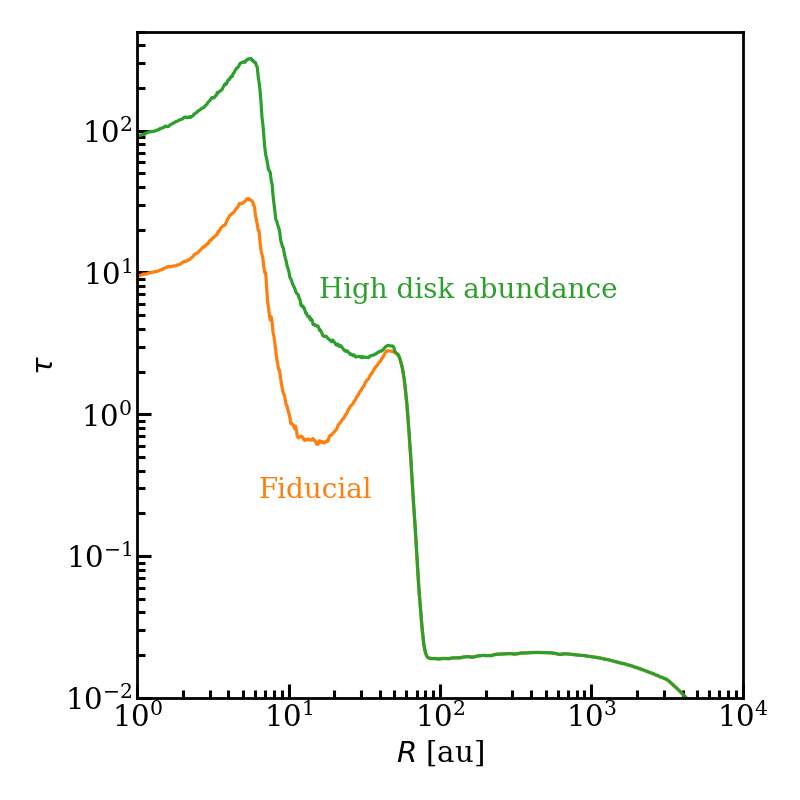}}
	\caption{Radial cut through the line optical depth of the image at the peak of the line for fiducial envelope-plus-disk model and the same with one order of magnitude higher disk methanol abundance.}
  \label{fig:tau_line_high_abund}
\end{figure}

\begin{figure}
	\resizebox{\hsize}{!}{\includegraphics{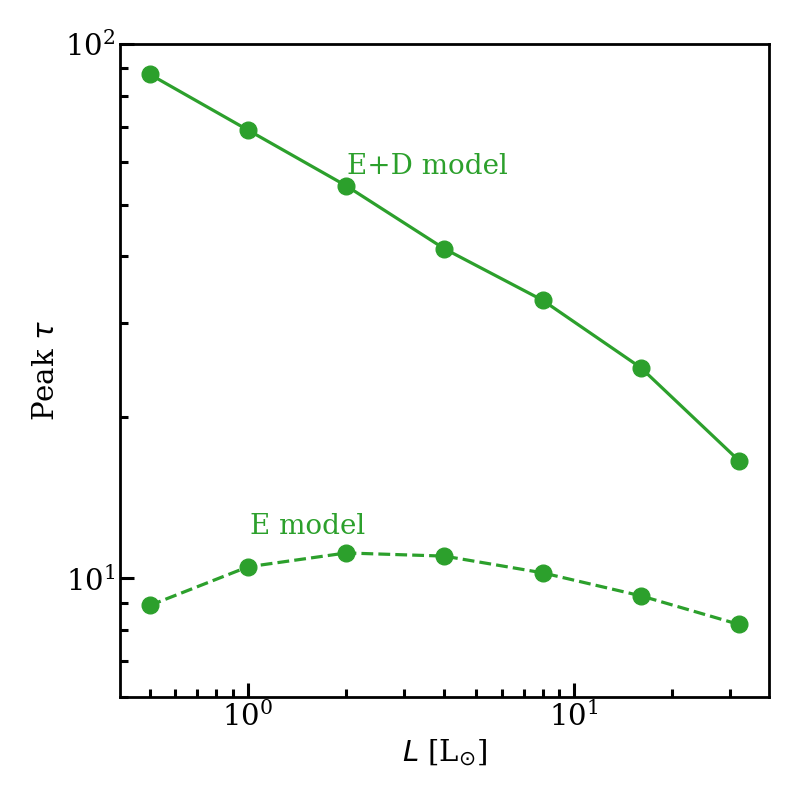}}
	\caption{The peak line optical depth of a radial cut for fiducial models with various luminosities.}
  \label{fig:tau_line_L}
\end{figure}

Figure \ref{fig:tau_line} shows the line optical depth for the fiducial models. The blue line in this figure is $\tau$ when the disk methanol abundance is set to zero so it shows $\tau$ in the envelope component of the fiducial envelope-plus-disk model. Figure \ref{fig:tau_line_M3} shows line optical depth as a function of radius for fiducial models with envelope mass of 3\,M$_{\odot}$ and the fiducial model with disk radius of 200\,au. Figure \ref{fig:tau_line_high_abund} shows the line optical depth versus radius for the fiducial envelope-plus-disk model and the same with one order of magnitude higher methanol abundance in its disk component. Figure \ref{fig:tau_line_L} shows the peak line optical depth of a radial cut for the fiducial models with different luminosities.

\begin{figure}
  \resizebox{\hsize}{!}{\includegraphics{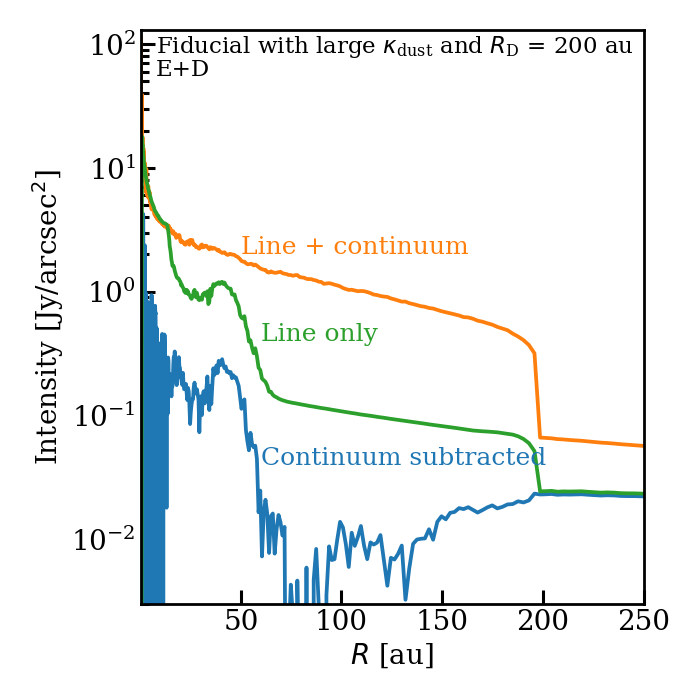}}
  \caption{The same as Fig. \ref{fig:continuum_oversub} but for the fiducial model with disk radius of 200\,au and high mm opacity dust grains.}
  \label{fig:continuum_oversub_200au}
\end{figure}

Figure \ref{fig:continuum_oversub_200au} shows the effect of continuum over-subtraction for the fiducial envelope-plus-disk model with high mm opacity dust grains and disk radius of 200\,au.

\section{Abundance of methanol}
\label{app:meth_abund}

In Eq. \eqref{eq:gas-grain} one needs to find the average of $a_{\rm d}^{2}$. Moreover, $n_{\rm d}$ can be written in terms of the average of $a_{\rm d}^{3}$. The average of $a_{\rm d}^{2}$ and $a_{\rm d}^{3}$ can be written as  

\begin{equation}
    {<}{a_{\rm d}^{2}}{>} = \frac{\int_{a_{\rm min}}^{a_{\rm max}} a^{2} a^{-\beta} da}{\int_{a_{\rm min}}^{a_{\rm max}} a^{-\beta} da} = \frac{1-\beta}{3-\beta} \times \frac{a_{\rm max}^{3-\beta} - a_{\rm min}^{3-\beta}}{a_{\rm max}^{1-\beta} - a_{\rm min}^{1-\beta}}, 
    \label{eq:a_squared}
\end{equation}
and
\begin{equation}
    {<}{a_{\rm d}^{3}}{>} = \frac{\int_{a_{\rm min}}^{a_{\rm max}} a^{3} a^{-\beta} da}{\int_{a_{\rm min}}^{a_{\rm max}} a^{-\beta} da} = \frac{1-\beta}{4-\beta} \times \frac{a_{\rm max}^{4-\beta} - a_{\rm min}^{4-\beta}}{a_{\rm max}^{1-\beta} - a_{\rm min}^{1-\beta}},
    \label{eq:a_cubed}
\end{equation}

\noindent respectively. Where $\beta$ depends on the dust distribution used and $a_{\rm min}$ and $a_{\rm max}$ are the minimum and maximum dust grain sizes in a dust distribution, respectively. In this work for the large dust grain distribution (high mm opacity dust) $\beta$ is assumed to be 3.5. The average of $a_{\rm d}^{2}$ for the high mm opacity dust distribution in this work is $1.25 \times 10^{-10}$\,mm$^{2}$ so that $\sqrt{{<}a_{\rm d}^{2}{>}} = 1.12 \times 10^{-5}$\,mm. 

Therefore, using Eqs. \eqref{eq:a_squared} and \eqref{eq:a_cubed}, the expression $\pi a_{\rm d}^{2} n_{\rm d}$ can be written in terms of hydrogen number density ($n_{\rm H}$), dust-to-gas mass ratio ($Q$), dust grains bulk density ($\rho_{\rm d}$), minimum and maximum dust grain sizes as the following 

\begin{equation}
     n_{\rm d}\pi a_{\rm d}^{2} = \frac{Q 1.4 m_{\rm p} n_{\rm H}}{\rho_{\rm d} 4/3 \pi {<}{a_{\rm d}^{3}}{>}} \pi {<}{a_{\rm d}^{2}}{>} = \frac{1.05 Q m_{\rm p} n_{\rm H} (4-\beta)}{\rho_{\rm d} (3-\beta)} \times \frac{a_{\rm max}^{3-\beta} - a_{\rm min}^{3-\beta}}{a_{\rm max}^{4-\beta} - a_{\rm min}^{4-\beta}}.
\end{equation}

\noindent Here $m_{\rm p}$ is the proton mass and factor 1.4 is to take into account for helium mass. Hence, for the small silicate dust grains used in this work where all the dust grains have the size of 0.1$\,\mu \rm m$ and $\rho_{\rm d} = 3.7\, \rm g\,cm^{-3}$, dust number density would be $1.5 \times 10^{-12} n_{\rm H}$.

\section{Critical density}
\label{app:crit_density}

\begin{figure}
  \resizebox{\hsize}{!}{\includegraphics{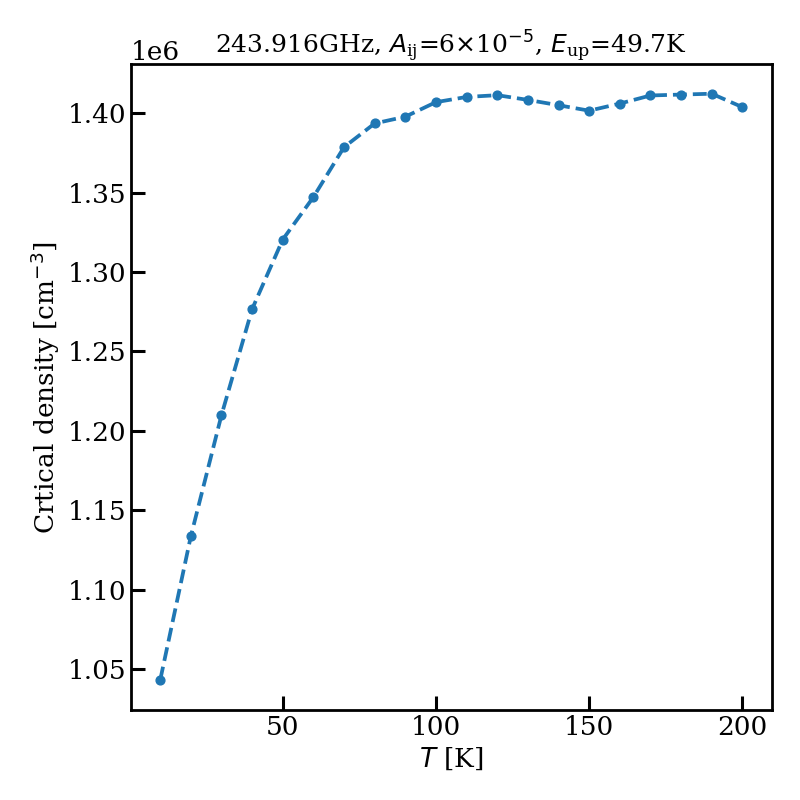}}
  \caption{The critical density for the methanol line studied in this work. The variation of critical density in this plot is less than a factor 2. }
  \label{fig:crit_dens}
\end{figure}

Figure \ref{fig:crit_dens} shows the critical density of the methanol line studied here as a function of temperature. The critical density is defined as the density above which collisional de-excitation of a specific level equals radiative de-excitation of that level. With this definition, one can conveniently estimate if the energy levels are populated according to a Boltzmann distribution. It is weakly dependent on temperature and computed for a particular level using the following equation (\citealt{Draine2011}),

\begin{equation}
    n_{\rm Crit}(T) = \frac{\sum_{l<u} A_{\rm u,l}}{\sum_{l<u} K_{\rm u,l}}.
    \label{eq:crit_dens}
\end{equation}

\noindent Where $A_{\rm u,l}$ is the Einstein $A$ coefficient for a transition with upper and lower levels u and l, respectively and $K_{\rm u,l}$ is the collision rate coefficient for that transition. The energy levels and Einstein $A$ coefficients for the line studied in this work were taken from the Cologne Database for Molecular Spectroscopy (CDMS) (\citealt{Muller2001, Muller2005}; \citealt{Xu2008}) and the collisional rate coefficients were taken from \cite{Rabli2010}.

\section{Grid of models}
\label{sec:grid}

\begin{figure*}
    \centering
    \includegraphics[width=0.69\textwidth]{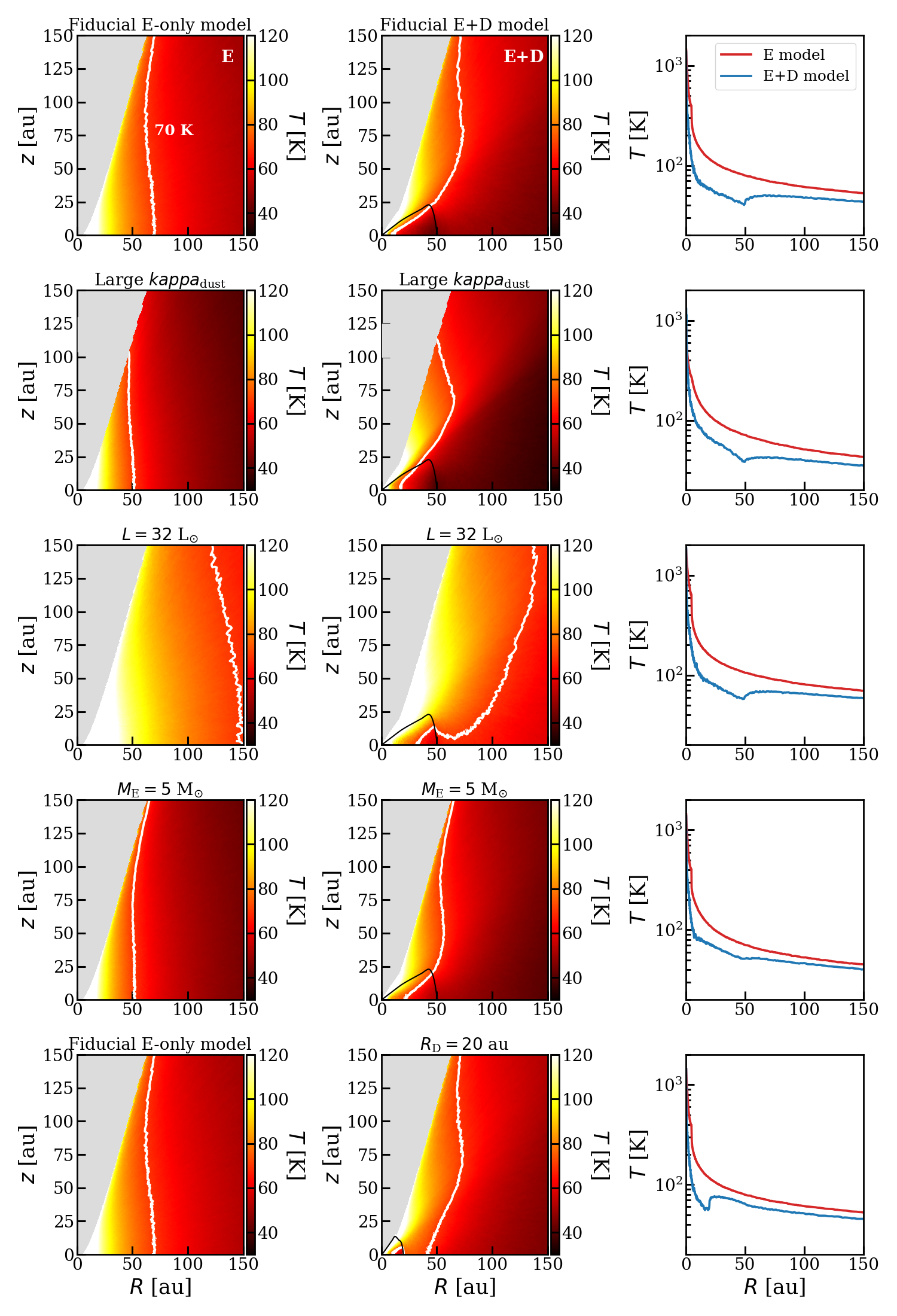}
    \caption{Two dimensional dust temperatures in the envelope-only models (left column) and in the envelope-plus-disk models (middle column). The white contours show the location where the temperature is 70\,K as an indication of the approximate temperature that methanol is sublimated at the densities of the models using the gas-grain balance model. The right column is the comparison of the mid-plane temperature in the envelope-plus-disk models and the envelope-only models. The rows from top to bottom present the fiducial model, the fiducial model when the dust grains have a high mm opacity, the fiducial model with stellar luminosity of 32\,L$_{\odot}$, the fiducial model when the envelope mass is 5\,M$_{\odot}$ and the fiducial model with disk radius of 20\,au, respectively. } 
    \label{fig:temp_grid}
\end{figure*}

Figure \ref{fig:temp_grid} shows the effects of various parameters of the disk and the envelope on the temperature structure.

\section{Additional plots}
\label{app:additional_plots}

\begin{figure}
  \resizebox{\hsize}{!}{\includegraphics{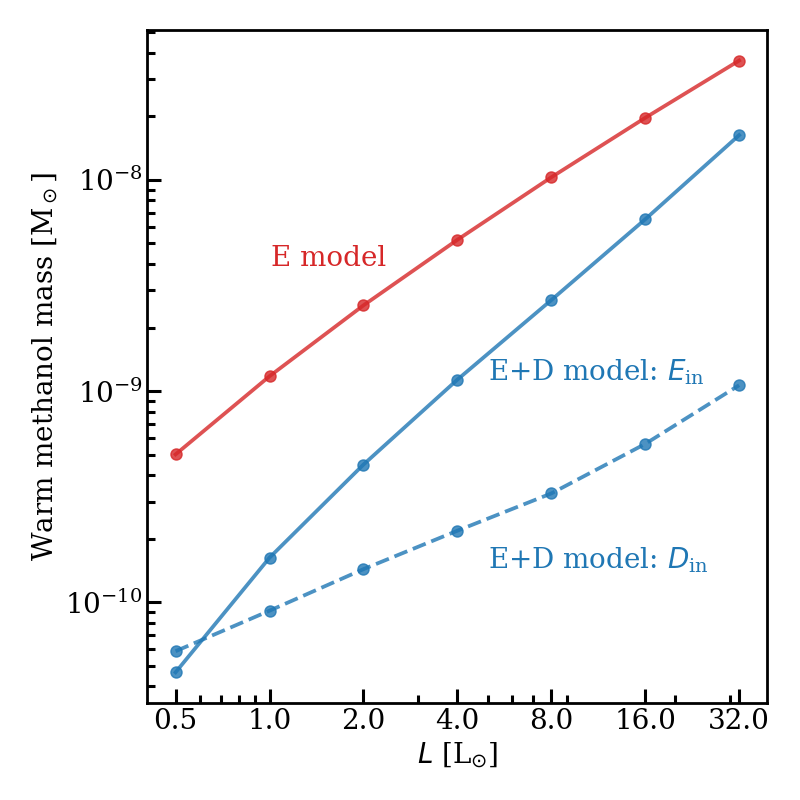}}
  \caption{The disk and envelope components of warm mass for the fiducial models with various luminosities. Red shows the envelope-only model and the blue shows the envelope-plus-disk model. The solid line shows the envelope-component (i.e. $E_{\rm in}$) of warm mass in the envelope-plus-disk model, whereas, the dashed line shows that for the disk component (i.e. $D_{\rm in}$).}
  \label{fig:hot_mass_components}
\end{figure}

\begin{figure}
  \resizebox{\hsize}{!}{\includegraphics{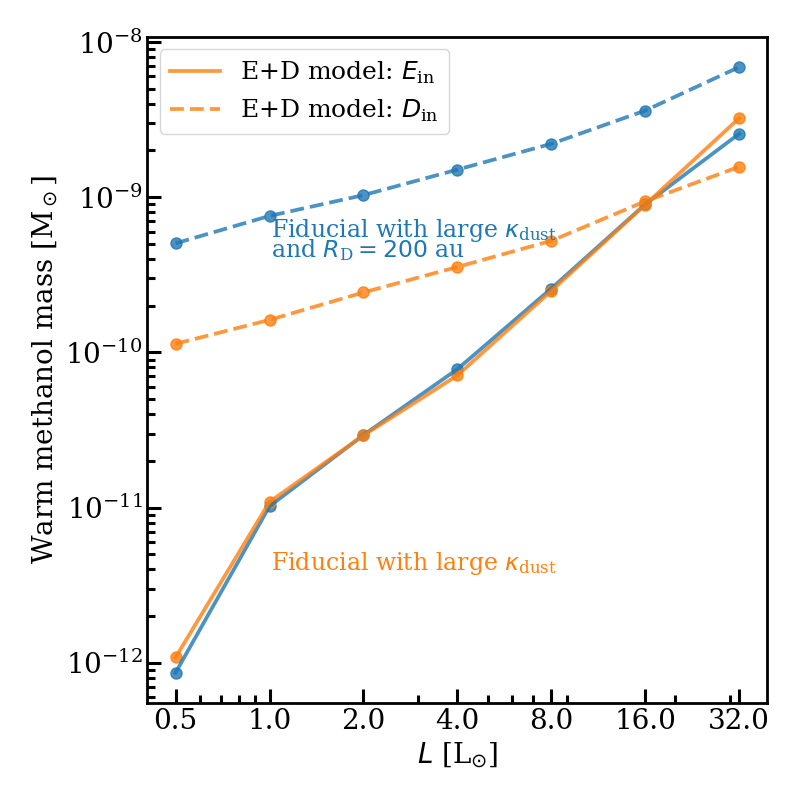}}
  \caption{Comparison between disk and envelope components of the fiducial model with high mm opacity dust grains and various luminosities (orange) and the same when disk radius is 200\,au. The solid line shows the envelope-component (i.e. $E_{\rm in}$) of warm mass in the envelope-plus-disk model, whereas, the dashed line shows that for the disk component (i.e. $D_{\rm in}$).}
  \label{fig:hot_mass_components_compare}
\end{figure}

Figure \ref{fig:hot_mass_components} shows the warm methanol mass for disk component and envelope component of the fiducial envelope-plus-disk model in blue and the same for the fiducial envelope-only model in red. This figure shows that the warm methanol mass is dominated by the envelope component of the envelope-plus-disk model. Figure \ref{fig:hot_mass_components_compare} shows the warm methanol mass in the disk and envelope components of the envelope-plus-disk models. Blue shows the fiducial model with high mm opacity dust grains and disk radius of 200\,au for various luminosities and orange shows the fiducial model with high mm opacity dust grains for various luminosities. There is more warm methanol mass in the disk component when the disk radius is 200\,au compared with when the disk radius is 50\,au. Moreover, for most luminosities shown here the warm mass is dominated by the disk component rather than the envelope component which is the opposite of the case of the fiducial model (low mm opacity dust grains, Fig. \ref{fig:hot_mass_components}). This is because of the larger photodissociation regions dominating the envelope component when the grains have a large mm opacity (Fig. \ref{fig:meth_abund} and \ref{fig:meth_abund_Rs_large}).

\begin{figure*}
    \centering
    \includegraphics[width=0.6\textwidth]{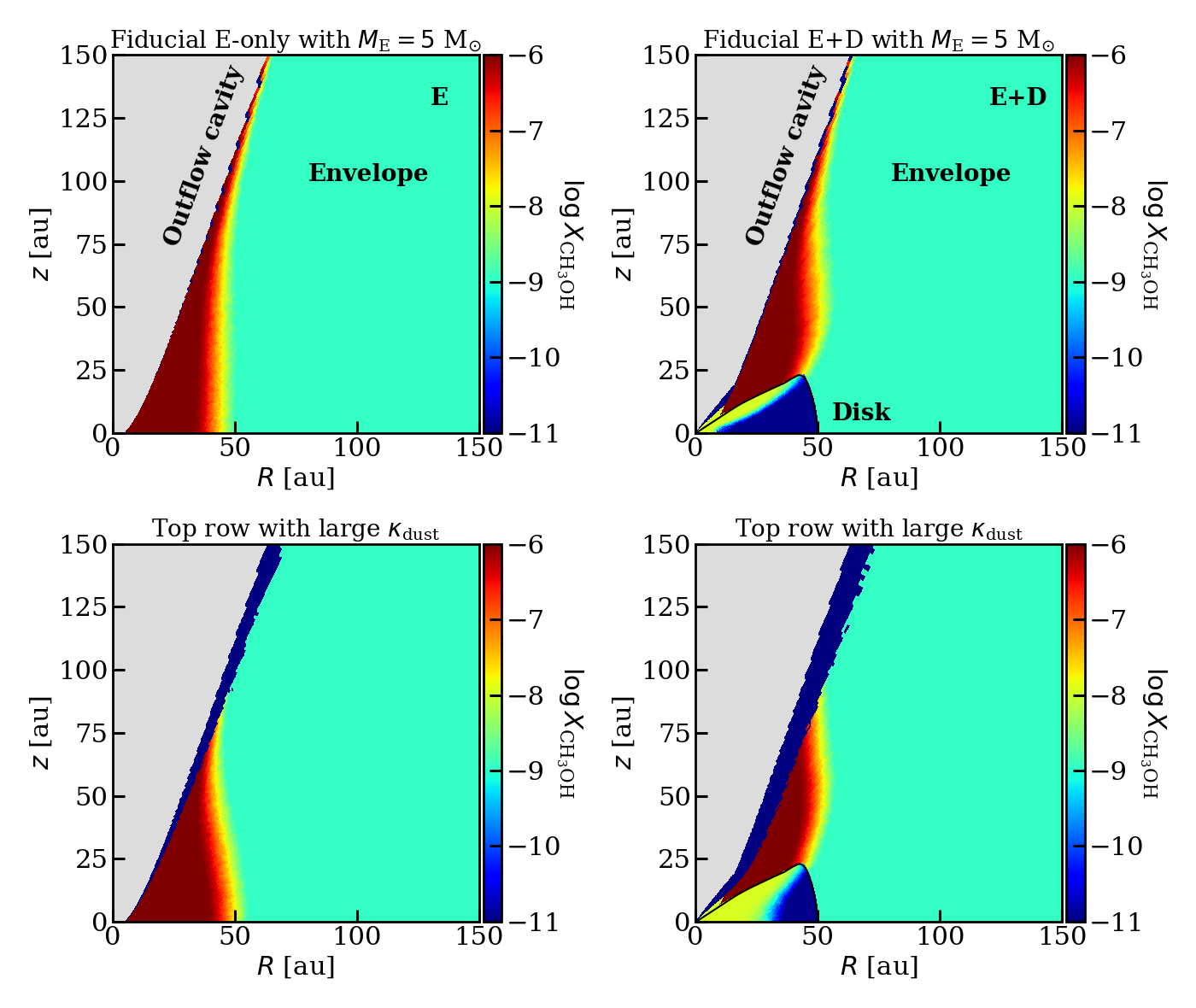}
    \caption{Methanol abundance maps for the envelope-only (left) and envelope-plus-disk (right) models when envelope mass is 5\,M$_{\odot}$ and other parameters are the same as the fiducial model. The top row shows the models with low mm opacity grains and the bottom row shows the models with high mm opacity grains. The photodissociation regions for $M_{\rm E}=5$\,M$_{\odot}$ models are very small and they are similar between models with low and high mm opacity grains which is different from the fiducial model seen in Fig. \ref{fig:meth_abund}.} 
    \label{fig:meth_abund_5M}
\end{figure*}

\begin{figure*}
    \centering
    \includegraphics[width=\textwidth]{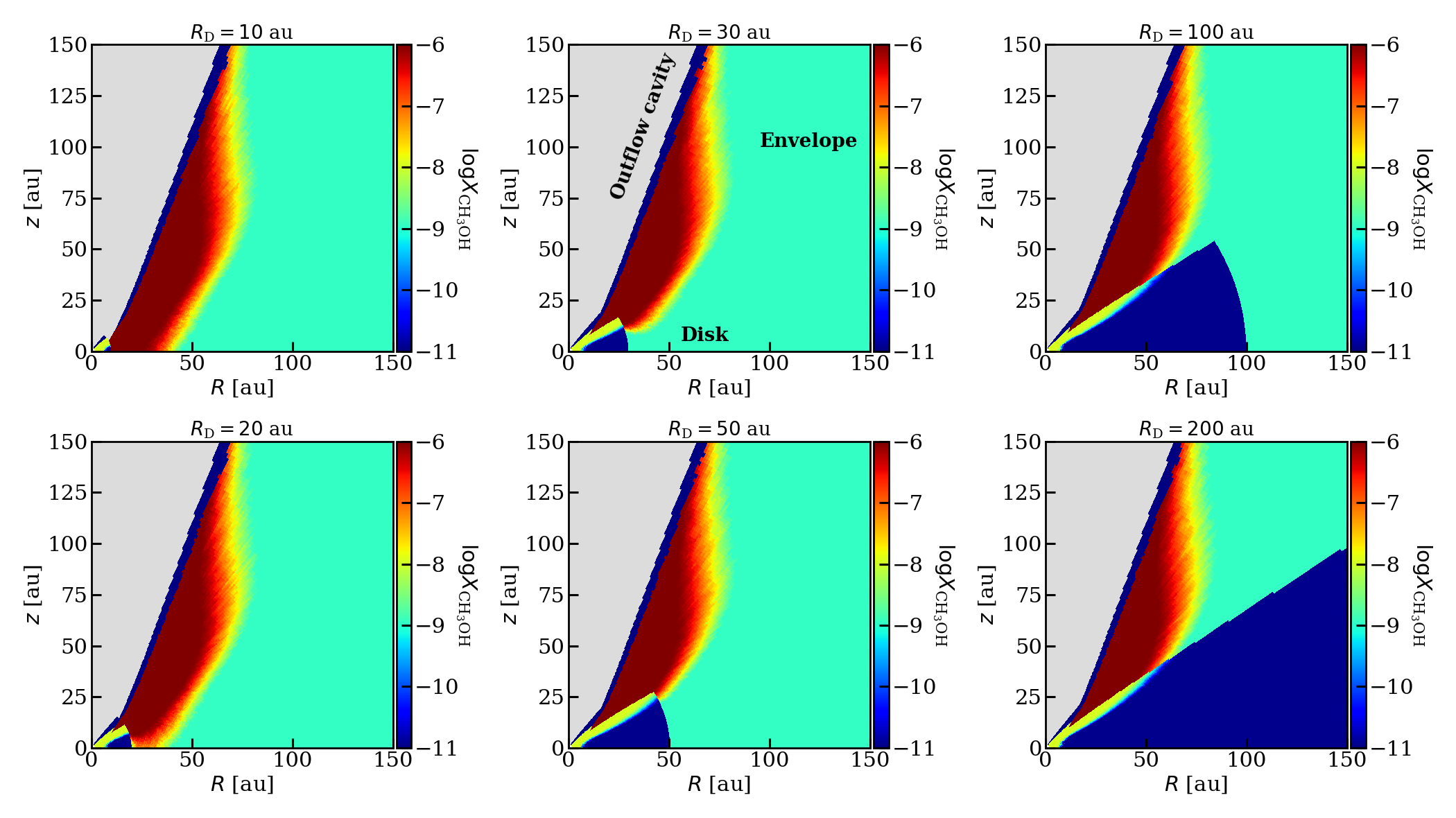}
      \caption{Methanol abundance maps for the fiducial envelope-plus-disk models with low mm opacity dust grains and different disk radii.} 
    \label{fig:meth_abund_Rs_small}
\end{figure*}

\begin{figure*}
    \centering
    \includegraphics[width=\textwidth]{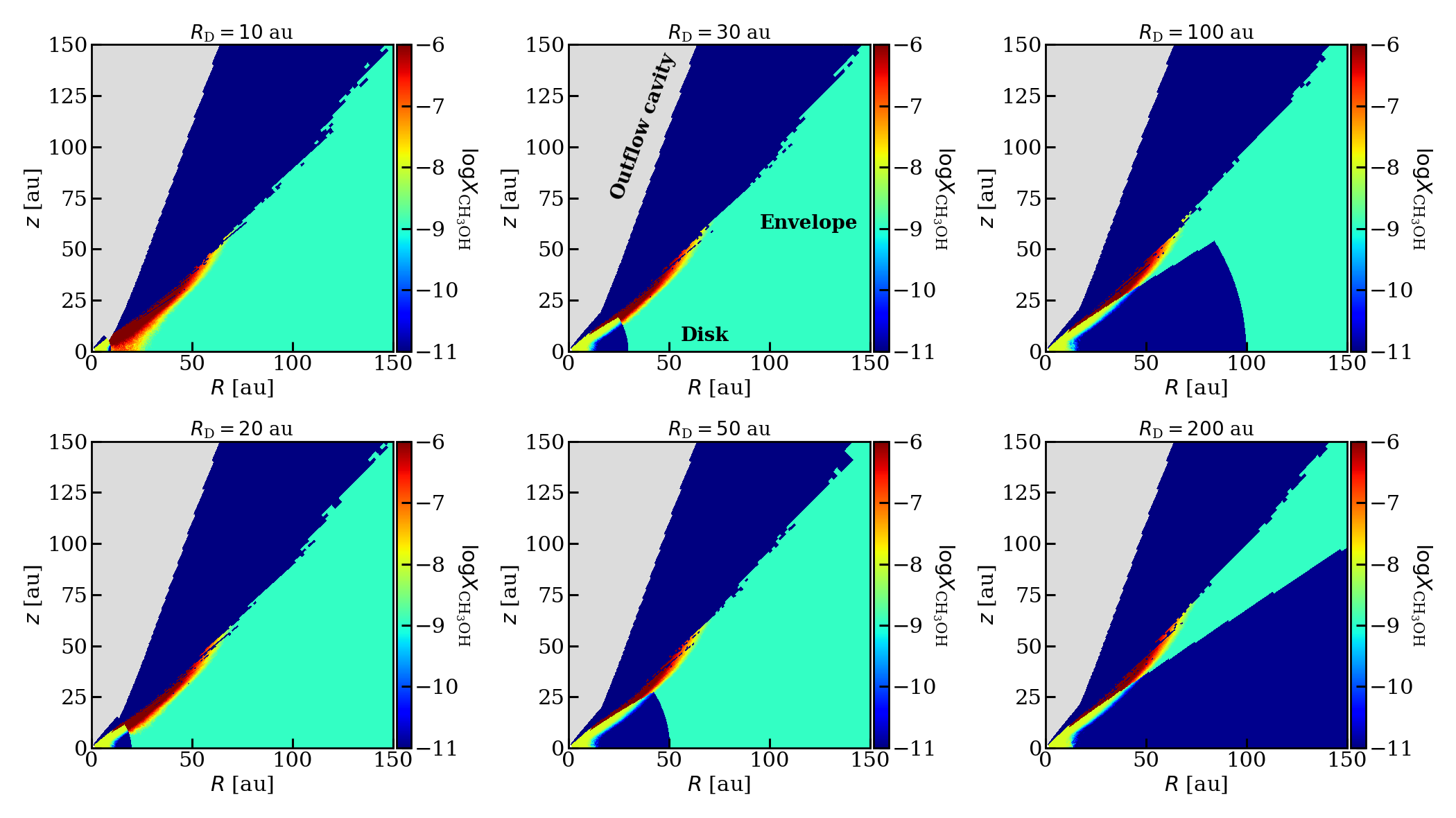}
      \caption{The same as Fig. \ref{fig:meth_abund_Rs_small} but now dust grains have a high mm opacity.} 
    \label{fig:meth_abund_Rs_large}
\end{figure*}

Figure \ref{fig:meth_abund_5M} shows methanol abundance maps for fiducial models with envelope mass of 5\,M$_{\odot}$. Figures \ref{fig:meth_abund_Rs_small} and \ref{fig:meth_abund_Rs_large} show methanol abundance maps for the fiducial envelope-plus-disk models with various disk radii for low and high mm opacity dust grains, respectively.

\begin{figure}
  \resizebox{\hsize}{!}{\includegraphics{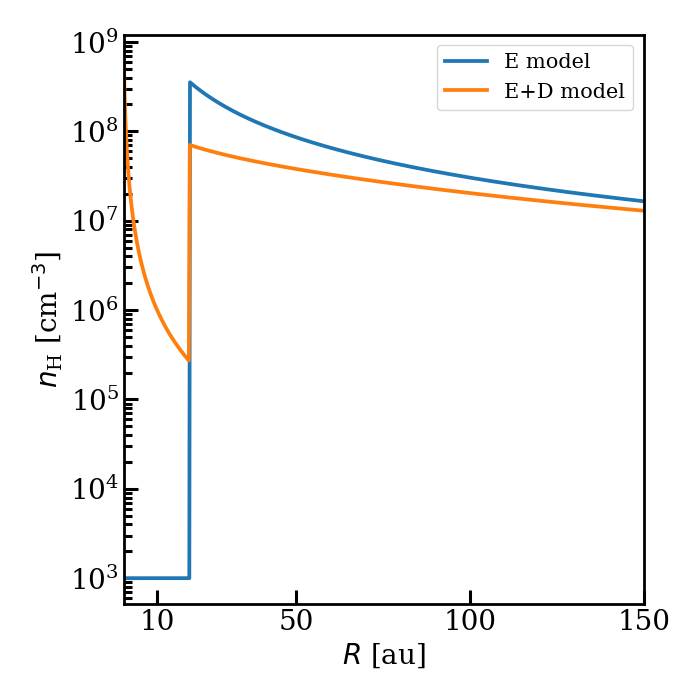}}
  \caption{A cut through the densities of the fiducial envelope-only and envelope-plus-disk model. The cut is in a direction just above where the disk is located to compare the densities in the envelope-only model and the envelope component of the envelope-plus-disk model.}
  \label{fig:dens_flattened}
\end{figure}

\begin{figure}
  \resizebox{\hsize}{!}{\includegraphics{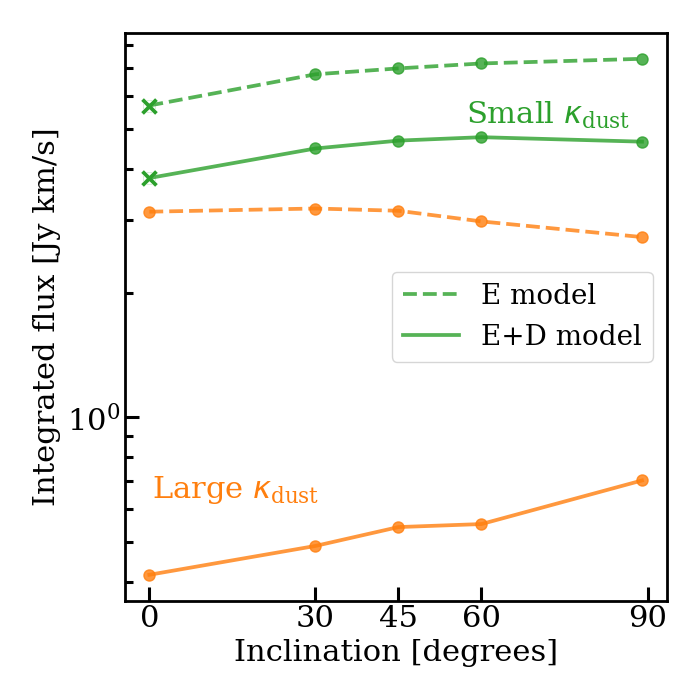}}
  \caption{The effect of viewing angle on the integrated line fluxes for the fiducial models.}
  \label{fig:angle_int}
\end{figure}

\begin{figure}
  \resizebox{\hsize}{!}{\includegraphics{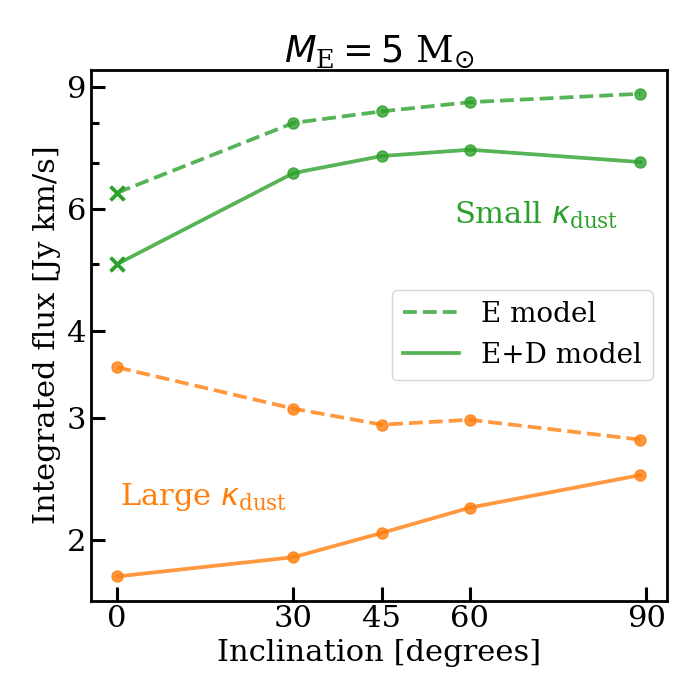}}
  \caption{The effect of viewing angle on the integrated line fluxes for the fiducial models with envelope mass of 5\,M$_{\odot}$.}
  \label{fig:angle_int_M5}
\end{figure}

Figure \ref{fig:dens_flattened} shows a cut through the density of total H for the fiducial envelope-only and envelope-plus-disk models to examine the difference between the density in the envelope component of the two models. Figure \ref{fig:angle_int} shows the effect of viewing angle on the integrated line fluxes for the fiducial models with low and high mm opacity dust grains. Figure \ref{fig:angle_int_M5} is the same as \ref{fig:angle_int} but for models with envelope mass of 5\,M$_{\odot}$.

\end{appendix}

\end{document}